\documentclass[journal]{vgtc}                

\ifpdf \usepackage[pdftex]{graphicx} \pdfcompresslevel=9
\else \usepackage[dvips]{graphicx} \fi

\usepackage{microtype}                 
\PassOptionsToPackage{warn}{textcomp}  
\usepackage{textcomp}                  
\usepackage{mathptmx}                  
\usepackage{times}                     
\usepackage{cite}                      
\usepackage{tabularx, ragged2e}
\newcolumntype{C}{>{\Centering\arraybackslash}X} 
\usepackage{color}
\usepackage{amsfonts}
\usepackage{paralist}
\usepackage{amsmath}
\usepackage{algorithm}
\usepackage[noend]{algpseudocode}
\usepackage{bbm}
\usepackage[utf8]{inputenc}
\usepackage[T1]{fontenc}
\usepackage{enumerate}
\usepackage{enumitem}
\usepackage{multirow}
\usepackage{bbold}
\usepackage{authblk}

\usepackage{amsmath}
\usepackage{autobreak}
\usepackage{wrapfig}

\usepackage[dvipsnames]{xcolor}
\usepackage[normalem]{ulem}
\definecolor{mred}{rgb}{.80,.12,.30}
\definecolor{MRED}{rgb}{.80,.12,.30}
\definecolor{grey}{rgb}{0.5,0.5,0.5}
\definecolor{lgrey}{rgb}{0.7,0.7,0.7}
\definecolor{purple}{rgb}{.75,0,.85}
\definecolor{pistachio}{rgb}{0.58, 0.77, 0.45}
\definecolor{myorange}{rgb}{0.94, 0.36, 0.13}

\newif\ifnotes
\notestrue

\let\origcite\cite
\renewcommand{\cite}[1]{\ifnotes\mbox{\origcite{#1}}\else \origcite{#1}\fi}

\onlineid{1088}

\vgtccategory{Research}
\vgtcpapertype{Empirical Study}

\title{Composition and Configuration Patterns in\\Multiple-View Visualizations}

\vgtcinsertpkg
\usepackage{graphicx}

\usepackage{subfigure}

\author{Xi Chen, Wei Zeng, Yanna Lin, Hayder Mahdi Al-maneea, Jonathan Roberts, and Remco Chang}
\authorfooter{
\item
Xi Chen, Wei Zeng, and Yanna Lin are with Shenzhen Institutes of Advanced Technology, Chinese Academy of Sciences.
Xi Chen is also with University of Chinese Academy of Sciences.
Wei Zeng is the corresponding author. E-mail: \{xi.chen2, wei.zeng, yn.lin\}@siat.ac.cn.

\item
Hayder M. Al-maneea and Jonathan C. Roberts are with Bangor University. E-mail: \{h.m.almaneea, j.c.roberts\}@bangor.ac.uk.
Hayder M. Al-maneea is also with University of Basrah.

\item
Remco Chang is with Tufts University. E-mail: remco@cs.tufts.edu
}

\abstract{
Multiple-view visualization (MV) is a layout design technique often employed to help users see a large number of data attributes and values in a single cohesive representation.
Because of its generalizability, the MV design has been widely adopted by the visualization community to help users examine and interact with large, complex, and high-dimensional data.
However, although ubiquitous, there has been little work to categorize and analyze MVs in order to better understand its design space.
As a result, there has been little to no guideline in how to use the MV design effectively.
In this paper, we present an in-depth study of how MVs are designed in practice.
We focus on two fundamental measures of multiple-view patterns: \emph{composition}, which quantifies what view types and how many are there; 
and \emph{configuration}, which characterizes spatial arrangement of view layouts in the display space.
We build a new dataset containing 360 images of MVs collected from IEEE VIS, EuroVis, and PacificVis publications 2011 to 2019, and make fine-grained annotations of view types and layouts for these visualization images.
From this data we conduct composition and configuration analyses using quantitative metrics of term frequency and layout topology.
We identify common practices around MVs, including relationship of view types, popular view layouts, and correlation between view types and layouts.
We combine the findings into a MV recommendation system, providing interactive tools to explore the design space, and support example-based design.
}
\begin{document}



\firstsection{Introduction}

\maketitle
We present an in-depth study on how multiple views are used in practice, and integrate our results into a recommendation system for the layout design of a multiple-view visualization.
Traditional visualization designs aim to maximize the utility of the visualization for specific data types or tasks. For example, line graphs show temporal information, maps display geographical information, etc.
Multiple-view visualization (denoted as MV) is a technique that seeks to integrate these visualizations by compositing multiple views of different view types into a single cohesive representation\cite{roberts_2007_state, javed_2012_exploring}.
Since each visualization conveys a specific perspective of data, a well-designed MV enables the user to simultaneously see representations of the same data from different perspectives.
In fact, the power of multiple views is well understood and the technique has nowadays become ubiquitous in exploratory data visualization\cite{roberts_2019_multiple}.

However, despite the ubiquity of multiple views in visualization systems, there are few guidelines, and those that do exist are very general.
For instance, researchers advise developers to use multiple views sparingly\cite{michelle_2000_guidelines}, and adopt consistent visual encodings across multiple views\cite{qu_2018_keeping}. 
Additionally, while researchers have made recommendations for multiple displays\cite{chung_2015_four}, and 
collaborative tasks over large displays\cite{langner_2019_multiple}, the plethora of design considerations pose challenges to developers in practice.

The visualization community has developed many visualization authoring tools, such as Power BI\cite{power_bi}, Tableau\cite{tableau}, and Spotfire\cite{spotfire}.
These tools allow the user to quickly design prototypes of MVs using a set of predetermined templates for common data types, such as the sales dashboard templates offered by Tableau\cite{tableau}.
However, for more complex data and tasks, designers often still need to manually curate the layout of MVs through trial and error.
This process can be tedious and time consuming, and sometimes produces results that fail to meet design guidelines\cite{qu_2018_keeping}.
Recently, researchers have developed techniques to automatically distribute multiple views in a visual space\cite{javed_2013_explates, sadana_2016_design, langner_2018_vistiles, horak_2019_vistribute}.
While the layout of theses systems may appear arbitrary, users place the views side-by-side in a deliberate way.

The goal of this paper is to create an image corpus of real-world MV images, analyze patterns contained in this data, distill a set of guidelines, and finally to produce a recommendation system for the design of MVs.
To code the design patterns of MVs, we first code each of the views in a MV in terms of its:

\begin{itemize}
\vspace{1mm}
\item
\emph{view type}: the mapping from data to visual form, \emph{i.e.}, the result of a visualization technique (\emph{e.g.}, bar and line charts);

\vspace{-1mm}
\item
\emph{bounding box}: position and size in the physical display space (most often in 2D) where the view is presented.
\end{itemize}

\vspace{-1mm}
After each view is coded in terms of its \emph{type} and \emph{bounding box}, we then encode the overall MV design based on its:

\begin{itemize}
\vspace{-1.5mm}
\item
\textit{Configuration}, including position and size of the bounding box of each view in the physical display space.

\vspace{-2mm}
\item
\textit{Composition}, including frequency, diversity, and correlations of view type usages. 
\end{itemize}
\vspace{-2mm}
Using this coding scheme, we collect and label images of MV designs from publications in \emph{IEEE VIS}, \emph{EuroVis}, and \emph{IEEE PacificVis} conferences from 2011 - 2019 (Section~\ref{ssec:mv_collection}).
The result is a curated dataset of 360 MV designs, which are then manually coded using an annotation tool that we developed for this effort (see Section~\ref{ssec:label_tool}).

We perform in-depth analyses on this dataset, using a number of quantitative metrics from information theory and graphics, such as conditional probability and layout topology.
The analyses reveal interesting composition and configuration patterns of MV design, including frequencies, aspect ratios, and positions of different view types (Section.~\ref{sec:analysis}).
For example, aspect ratios of most view types are within [1/2, 2], except for some types like \emph{Area} and \emph{Panel} (see Figure~\ref{fig:viewTypeRatio}).

Lastly, using the found composition and configuration patterns from the analyses, we develop an interactive recommendation system for designing MVs. In particular, this system:
(1) enables multi-faceted exploration of existing MV designs (Section~\ref{ssec:exp_mode}), and (2) recommends designs given view type and layout information (Section~\ref{ssec:design_mode}).
We evaluate the utility and effectiveness of this recommendation via a formal user study.
The results of the study suggest that our recommendation tool can enhance the understanding of MV design space, and promote appropriate MV designs, for both visualization novices and experts.

\vspace{1mm}
In summary, we contribute to the MV design in the following way:
\begin{itemize}

\vspace{-2mm}
\item
We curate a new dataset of 360 MV designs. Using an annotation tool developed for this project, each of the MV designs has been carefully labeled and annotated. The resulting dataset will be released publicly to foster future research.

\vspace{-2mm}
\item
We conduct a series of quantitative analyses on the dataset, and find common composition patterns of view type usage and configuration patterns of layout arrangement in the design of MVs.

\vspace{-2mm}
\item
We develop an interactive recommendation system that supports multi-faceted exploration, and recommendations of MV designs.
We conduct a formal user study showing the effectiveness of the recommendation system.
The system is freely available for the academic purpose at~\url{https://mvlandscape.bitbucket.io/}.
\end{itemize}
\section{Related Work}
\label{sec:related_work}

\textbf{Multiple Views}.
Card et al.'s reference model\cite{card_1999_readings} states that visualizations are created in four steps:
i) processing \emph{raw data} into \emph{data tables}; 
ii) mapping \emph{data tables} to \emph{visual structures};
iii) transforming \emph{visual structures} to \emph{views} through operations like zooming and brushing;
and iv) rendering and displaying \emph{views} to users.
While the reference model is useful for designing a single visualization, it does not provide guidelines for designing visualizations for more complex and high-dimensional data.
To help users examine and interact with large, complex, and high-dimensional data, multiple-view visualizations (MV) that can show different perspectives of data emerged\cite{roberts_1998_encouraging}.
For example, dashboards evolve from single- to multiple-view visualizations, rendering an increase of data visibility, enhancement of operational efficiency, and reduction of understanding cost\cite{sarikaya_2019_what}.
The visualization community has contributed to MV design from various perspectives, \emph{e.g.}, suggesting rules and guidelines\cite{michelle_2000_guidelines, qu_2018_keeping}, developing authoring tools (\emph{e.g.}, Polaris\cite{stolte_2002_polaris}, Improvise\cite{weaver_2004_improvise}, and ComVis\cite{matkovic_2008_comvis}), and extending to mobile devices\cite{sadana_2016_design, langner_2018_vistiles, horak_2019_vistribute} and large displays\cite{langner_2019_multiple}.

Many theories have also been proposed to facilitate the understanding of the relationship between views.
For example, VisLink\cite{collins_2007_vislink} formalized multi-relation visualizations as side by side, in parallel, or in chosen placements.
Javed and Elmqvist\cite{javed_2012_exploring} and Gleicher et al.\cite{gleicher_2011_visual} categorized design space of composite visualizations into juxtaposition, superimposition, overloading, nesting, and integration.
However, though much progress has been made, the design of the layout of a multiple views visualization is usually curated manually based on the designers' experience.
This process can be difficult and laborious for professional designers, not to mention visualization novices.

The difficulty in designing MVs suggests a lack of structure and understanding of the design space of MVs.
However, the view layouts are still created by humans, which suggests that MV design is not arbitrary\cite{horak_2019_vistribute}.
This work helps to address this challenge, by performing an empirical study on how MVs are designed in practice, which we categorize as \emph{composition} and \emph{configuration} patterns.
The goal of this study is to further our understanding of the design space of MVs and provide the foundation for data-driven MV design.

\vspace{1.5mm}
\noindent
\textbf{Data-driven visualization design}. 
Mackinlay proposed APT (A Presentation Tool)\cite{mackinlay_1986_automating} for automated visualization design based on the expressiveness and effectiveness of the visualization.
APT builds on studies in graphical perception, \emph{e.g.}, rankings of visual variables by data type\cite{cleveland_1984_graphical} and analytical tasks\cite{casner_1991_task}.
Some of these findings have been integrated into the development of visualization authoring tools, \emph{e.g.}, ShowMe\cite{mackinlay_2007_tabulea}.
Following Mackinlay's work, there has been an increasing trend of using data-driven models for automated visualization design\cite{saket_2018_beyond}.
Most of these studies aim to learn an optimal mapping from inputs of data attributes and tasks to outputs of visualizations.
For example, SEEDB\cite{vartak_2015_seedb} recommends visualizations that it deems useful or interesting based on the perceived utility of the visualization;
Data2Vis\cite{dibia_2019_data2vis} learns an end-to-end model for automatic visualization generation; DeepEye\cite{luo_2018_deepeye} and VizML\cite{hu_2019_vizml} learn to rank visualization for input specifications of data, tasks, and context; Draco\cite{moritz_2019_draco} learns soft constraints for visualization design; and Chen et al.\cite{chen_2019_towards} learns global and local features for timeline infographics.
While useful, these works focus on mapping \emph{data tables} to \emph{visual structures} for a single visualization.
The goal of this paper is to provide the foundation for designing the composition and configuration of these visualizations into a multiple-view design.

In the simplest form of a MV design, individual visualizations can be arranged in a grid layout, as shown in VizDeck\cite{alicia_2012_vizdeck}.
However, as we show in Section~\ref{sec:analysis}, real-world composition and configuration patterns of MVs can be more complicated and divergent.
Due the complexity of the design space, we develop a recommendation system to help a designer generate a MV design.
Based on the analyses of the MVs in the VIS community over the last nine years, our recommendation system can propose useful and more nuanced MV designs that are more suited for real-world applications.

\vspace{1.5mm}
\noindent
\textbf{Layout Design}.
MV design can be regarded as a layout design problem studied in many fields, such as graphics design (\emph{e.g.},\cite{xu_2014_global, donovan_2014_learning}), architecture (\emph{e.g.},\cite{yang_2013_urban, wu_2018_miqp}), and treemap (\emph{e.g.},\cite{shneiderman_1992_tree, bederson_2003_ordered}).
Here we briefly summarize closely related works in graphics design and treemap.

Many works in graphic layout design largely rely on rule-based approaches based on existing design principles.
For example, Xu et al.\cite{xu_2014_global} introduced the beautification metric for the global layout that aligns sketch-based interfaces.
However, the design principles typically aim to optimize certain properties, which may ignore the resulting effect on other aspects of the design.
As a result, in recent years researchers have begun to explore an exemplar-based approach by learning from existing designs.
For example, O’Donovan et al.\cite{donovan_2014_learning} optimized the arrangement of the input contents of a single-page infographic layout based on a small number of example layouts.
Zheng et al.\cite{zheng_2019_content-aware} showed that infographic layout can be synthesized using a deep generative model learned from a large-scale magazine layout dataset.

Related to the layout problem in graphic design, treemap is a nested enclosure visualization, which is often described as space-filling\cite{shneiderman_1992_tree}.
The design of treemaps (e.g. squarified treemap\cite{Bruls99squarifiedtreemaps}) therefore shares similar considerations as the design of MVs.
In particular, an effective MV design should also take into account the efficiency of the usage of space, which can be evaluated by quantitative metrics proposed for measuring the space efficiency of treemaps.
Borrowing from literature in treemap design, we adopt a series of metrics, including \emph{aspect ratio}\cite{bederson_2003_ordered}, and \emph{stability} and \emph{relative-position-change}\cite{sondag_2018_stable} when analyzing the configuration pattern of MVs.
\section{Definition and View Specification}
\label{sec:overview}

\begin{figure}[t]
  \centering
  \includegraphics[width=0.485\textwidth]{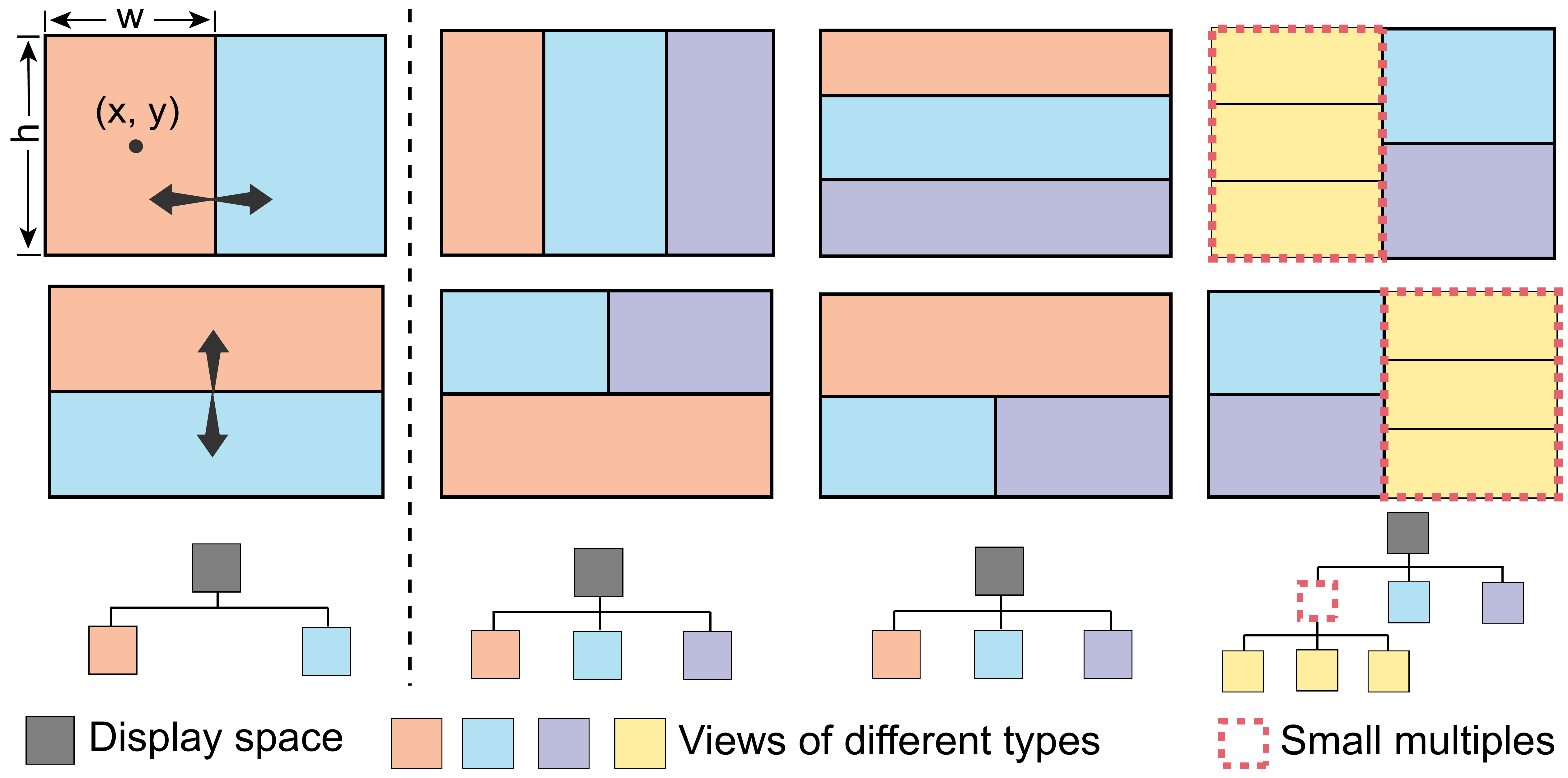} \\
  \vspace*{-4mm}
  \caption{Illustration of some example MV layouts.
  Tree structures below depict corresponding hierarchies of the views.}
  \label{fig:mv_model}
  \vspace*{-5mm}
\end{figure}

We define a multiple view (MV) as a layout that arranges two or more views in a display space (see Figure~\ref{fig:mv_model} for some example layouts).
Each view in a MV consists of at least two perspective attributes:

\begin{itemize}[leftmargin=*]

\vspace{-2mm}
\item
\textit{View type} (denoted as \emph{type}):
Following Card et al.'s model\cite{card_1999_readings}, a view is formed by applying transforming operations to the visual structure.
Many visual structures have been designed,
which can be classified into various view types\cite{gerald_1994_classification, schneiderman_1996_eyes}.
Recently, Borkin et al.\cite{borkin2013makes} created a taxonomy of 12 chart types for information visualization, including \emph{Bar}, \emph{Line}, and \emph{Circle}, etc.
We adopt the taxonomy in this study.
In addition, we refer to scientific visualizations, including volume and flow visualization, collectively as \emph{SciVis}.
We treat graphic widgets, including menus, legends, and narrative texts that are not overlaid on top of other views, as \emph{Panel}.
In practice, many widgets are arranged side-by-side, in which cases we treat them as one panel.
Sometimes a MV arranges several menus on the periphery, forming a very narrow region that makes a marginal effect on the view layout.
We ignore these menus following the convention established in\cite{al-maneea_2019_multipleview}.

In summary, in this work, we consider 14 views types (12 types of information visualizations\cite{borkin2013makes} $+$ SciVis $+$ Panel).
The use of the view types defines the \emph{composition} pattern of a MV.
The view types, together with their abbreviations and icon representations are presented in Supplementary Table S1, and examples of each view type are presented in Supplementary Table S2.

\vspace{-2mm}
\item
\textit{Bounding box} (denoted as \emph{bbox}) specifies the center position $(x, y)$ and size $w \times h$ of a view in the display space $S$.
Multiple boxes fill up the display space, \emph{i.e.}, no gaps or overlaps between two neighboring views.
The display space $S$ can fall in a wide range from small-size tablets, to medium-size desktop monitors, to large-size video walls.
However, MVs studied in this work are collected from publication figures that do not state exact display size.
Thus we adopt normalized parameters for \emph{bbox}, \emph{i.e.}, $x, y, w, h$ are scaled to the range of [0, 1].

The arrangement of multiple \emph{bbox}es in the display space formalizes the view layout, \emph{i.e.}, \emph{configuration} pattern of a MV. 
\end{itemize}

\vspace{-2mm}
In practice, many MVs juxtapose several views of the same visual type together. 
This is referred to as ``small multiples'' by Tufte\cite{tufte_1983_visual}.
To distinguish small multiples from other MV designs, we further incorporate the concept of hierarchy:
the display space $S$ is regarded as the root node;
views and small multiples filling up $S$ are level-1 nodes;
views forming level-1 small multiples are level-2 nodes; and so on.
Although the process can repeat multiple times, we find that views in our corpora are at most level-2.
Figure~\ref{fig:mv_model} illustrates some example layouts, and the tree structures below depict the view hierarchies.
MVs on the right side are made up of five views, three of which (in yellow color) form a small multiples. 

Given the constraint of a 2-level hierarchy, we can define a MV as:
\vspace{-2mm}
\begin{center}
$
MV := \{view_1, [view_2, view_3],\cdots,view_n\}
$
\end{center}
\vspace{-2mm}
where $view_i$ is represented as a two-tuple $view_i := (type_i, bbox_i)$, and $[\cdots]$ denotes a small multiples that contains several level-2 views. 
\section{Dataset Construction}
\label{sec:data}

\subsection{Multiple-View Visualization Collection}
\label{ssec:mv_collection}

\vspace{-2mm}
\begin{table}[h]
\caption{Number of MVs collected in IEEE VIS (VAST, InfoVis, SciVis), EuroVis, and IEEE PacificVis from 2011 to 2019.}
\begin{tabular}{c}
\begin{minipage}{0.495\textwidth}
\includegraphics[width=0.95\textwidth]{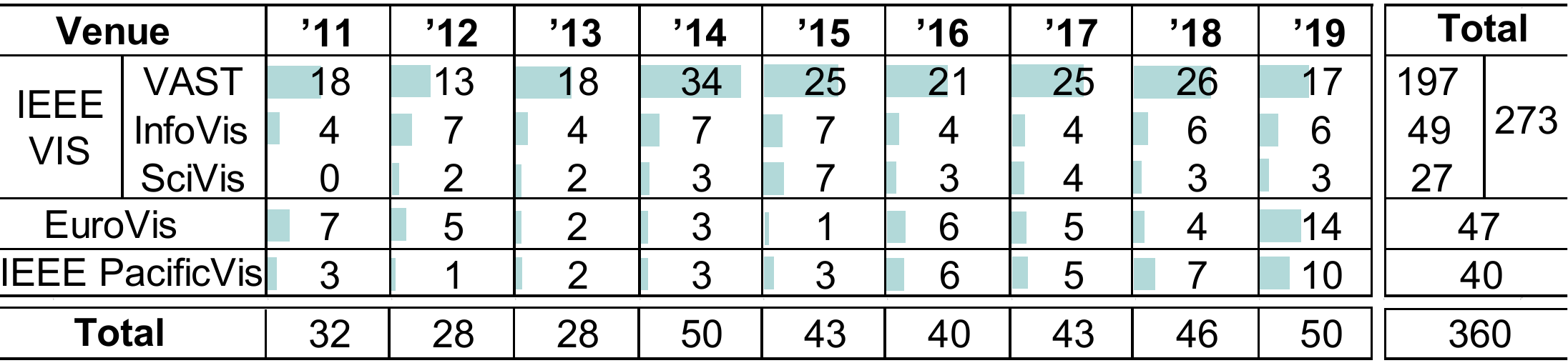}
\end{minipage}
\end{tabular}
\label{tab:conf-num}
\vspace{-3.5mm}
\end{table}

To ensure design diversity and quality, we select MVs from publications in \emph{IEEE VIS}, \emph{EuroVis}, and \emph{IEEE PacificVis} from 2011 - 2019.
The MVs are initially recorded as images, which are collected through the following steps:

\begin{itemize}
\vspace{-2mm}
\item
First, we crawl the publications by referring to their digital object identifiers (DOIs), using the information collected in the dataset\cite{isenberg_2017_vispubdata} for \emph{IEEE VIS} publications, and from the DBLP database for the other publications.
The process produces a total of 1,976 publications, including 1,149 from \emph{IEEE VIS}, 475 from \emph{EuroVis}, and 352 from \emph{IEEE PacificVis}.

\vspace{-2mm}
\item
Next, we use a combination of several data preparation and image processing techniques to automatically extract figures from the papers:
(1) We employ \textit{PyMuPDF}\footnote{www.pypi.org/project/PyMuPDF/} and \textit{pdftohtml}\footnote{www.sourceforge.net/projects/pdftohtml/} to convert \emph{pdf} papers to \textit{jpg} and \textit{xml} files, respectively;
(2) By querying the keywords of \emph{Fig.} or \emph{Figure} in the \textit{xml} file, we can locate positions of all figures in one paper; 
and (3) We crop all figures from the \textit{jpg} file based on the figure positions.
In this way, we extract 16,891 figures from 1,976 papers. 

\vspace{-2mm}
\item
Lastly, we manually choose MVs made up of two or more views, by filtering out figures that fall into one of the following conditions:
(1) figures of system interfaces, \emph{e.g.}, architecture diagrams;
(2) figures that only include a single view visualization;
and (3) figures that contain multiple interfaces.
These conditions filter out most figures, yet there still exist some figures that are difficult to verify from the image.
In such cases, we check carefully the figure captions and corresponding text descriptions in the papers.
Finally, if there is more than one MV figure in a paper, we choose the first one that appears in the paper.
\end{itemize}
\vspace{-1mm}

In this way, we identify a total of 360 MV images. 
Table~\ref{tab:conf-num} presents the number of MVs collected in each conference per year.
Most MVs are collected from \emph{IEEE VIS} since there are many more papers in \emph{IEEE VIS} than \emph{EuroVis} and \emph{PacificVis}.
In detail, the table shows that \emph{IEEE VAST} contributes the most (197/360) since systems in VAST typically utilize the MV design to present complex datasets.
It is interesting to notice that \emph{IEEE SciVis} has the smallest number.
This is probably because many SciVis papers present advanced algorithms in rendering or interaction, which do not require multiple views.
We also notice that numbers of MVs in \emph{EuroVis} and \emph{PacificVis} increased much in 2019, and the number of MVs in all the conferences is slightly growing.

\begin{figure}[h]
  \centering
  \includegraphics[width=0.495\textwidth]{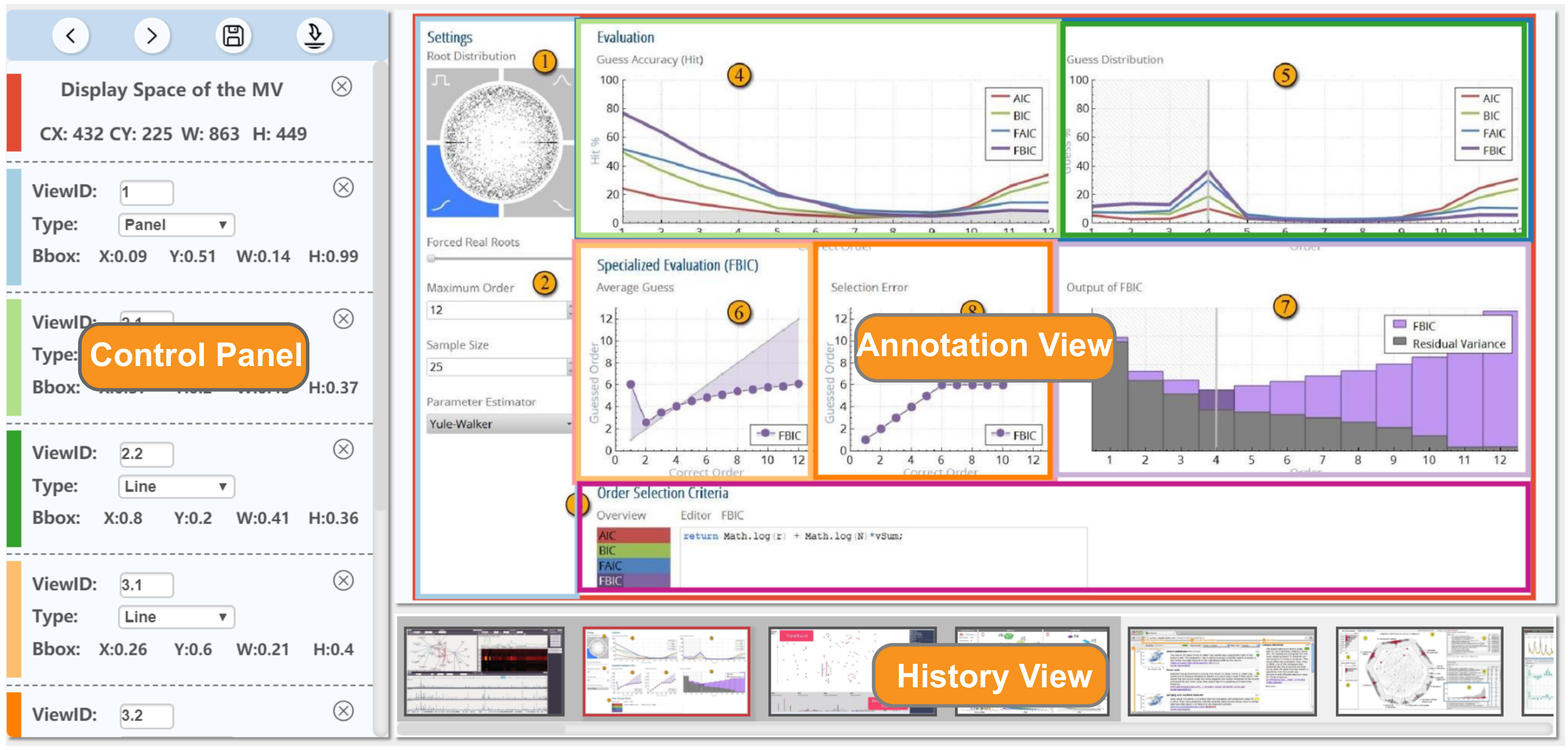}
  \vspace*{-6mm}
  \caption{Our annotation tool consists of three components: Annotation View, Control Panel, and Historical View.}
  \label{fig:label_tool}
  \vspace*{-5mm}
\end{figure}

\subsection{View Annotation}
\label{ssec:label_tool}

After collecting images of MV designs, we need to extract attributes of \emph{type} and \emph{bbox} for each view.
First, we tried automated approaches, such as deep learning techniques.
We trained a YOLOv3\cite{yolov3} model using visualization images self-generated and from open datasets such as\cite{hu_2019_viznet}.
We applied the model to predict views in the MV images, which however, produced unsatisfactory results.
For instance, the intersection over union (IOU) $-$ a standard performance metric that measures the size of intersection divided by the size of union for one pair of bounding boxes, was 73.3\% for views of parallel coordinates.
This accuracy could not support further analysis, so we decided to manually label the views.

We opted to develop an annotation tool that is dedicated to label attributes of \emph{type} and \emph{bbox} of each view.
As shown in Figure~\ref{fig:label_tool}, our annotation tool consists of three components. 
\begin{itemize}
\vspace{-2mm}
\item
\textbf{Annotation View} is the main view that displays the MV image being annotated.
The view allows users to annotate rectangles within the image $I \in \mathbb{R}^{W_I \times H_I}$, where $W_I$ \& $H_I$ indicate width and height of the MV image.
First, users need to annotate a rectangle that specifies display space $S \in \mathbb{R}^{W_S \times H_S}$ of the interface, where $W_S \leq W_I$ \& $H_S \leq H_I$.
Next, users can draw rectangles within $S$.
Each rectangle indicates $bbox_i$ for a view $view_i$, and is assigned to a unique color.
We constrain $bbox_i$ within $S$ by clipping the intersection of user-specified rectangle $rect_i$ and $S$, \emph{i.e.}, $ bbox_i = rect_i \cap S$.
Users can reposition and resize a $bbox_i$ by dragging control points on the corner of each edge.
We allow overlap or gap between two neighboring $bbox_i$ and $bbox_j$, which will be fixed in a post-processing stage (see Section~\ref{ssec:alignment}).

\vspace{-2mm}
\item
\textbf{Control Panel} consists of a set of widgets allowing users to navigate or change view attributes.
In the top, the blue region lays four buttons of \emph{previous}, \emph{next}, \emph{save}, and \emph{load}.
Users can view the previous/next interface by clicking on \emph{previous} or \emph{next} button, respectively.
If users feel comfortable with current annotations, they can \emph{save} them; if users would like to make changes, they can \emph{load} previous annotations.

When a user annotates a rectangle in the \textit{Annotation View}, a new section with a text field of \emph{View ID} and a drop list of \emph{Type} will be added to the panel.
Each section is marked in the same color with the corresponding rectangle color in the \textit{Annotation View}.
In the \emph{View ID} text field, the user can input a positive integer (\emph{e.g.}, 1, 2) indicating the view is a level-1 view, or a one-decimal number (\emph{e.g.}, 3.1, 3.2) indicating the view is a level-2 view, where the integral part indicates the small multiples ID, and the decimal part indicates the view number within the small multiples.
In the \emph{Type} drop list, the user can choose one out of the 14 view types (see Section~\ref{sec:overview}).
Users can also choose to delete the view using the \emph{trash} button attached besides.

\vspace{-2mm}
\item 
\textbf{History View} overviews all MVs in the dataset by showing an image thumbnail for each MV.
Annotated MVs are shown in the gray background; the one being annotated is marked with a red outline, and unannotated ones are shown in the white background.
\end{itemize}
\vspace{-1mm}

After finishing annotating a MV, we store the labeling results in a JSON file named by the paper DOI.
The JSON file records center position and size of the MV in the image, and an array named \textit{views} storing information of \emph{type} and \emph{bbox} of all views. 
Level-2 views in a small multiples are stored as a nested array named \textit{small multiples} within \emph{views} array.
Specifically, $bbox_i$ of a view $view_i$ stores the normalized center position $(x_i, y_i)$ and size $w_i \times h_i$ to facilitate comparison among different MVs.
By referring to the interface position and size, we can recover the exact position of each view in the image.

\subsection{Layout Refinement}
\label{ssec:alignment}

\begin{figure}[t]
  \centering
  \includegraphics[width=0.495\textwidth]{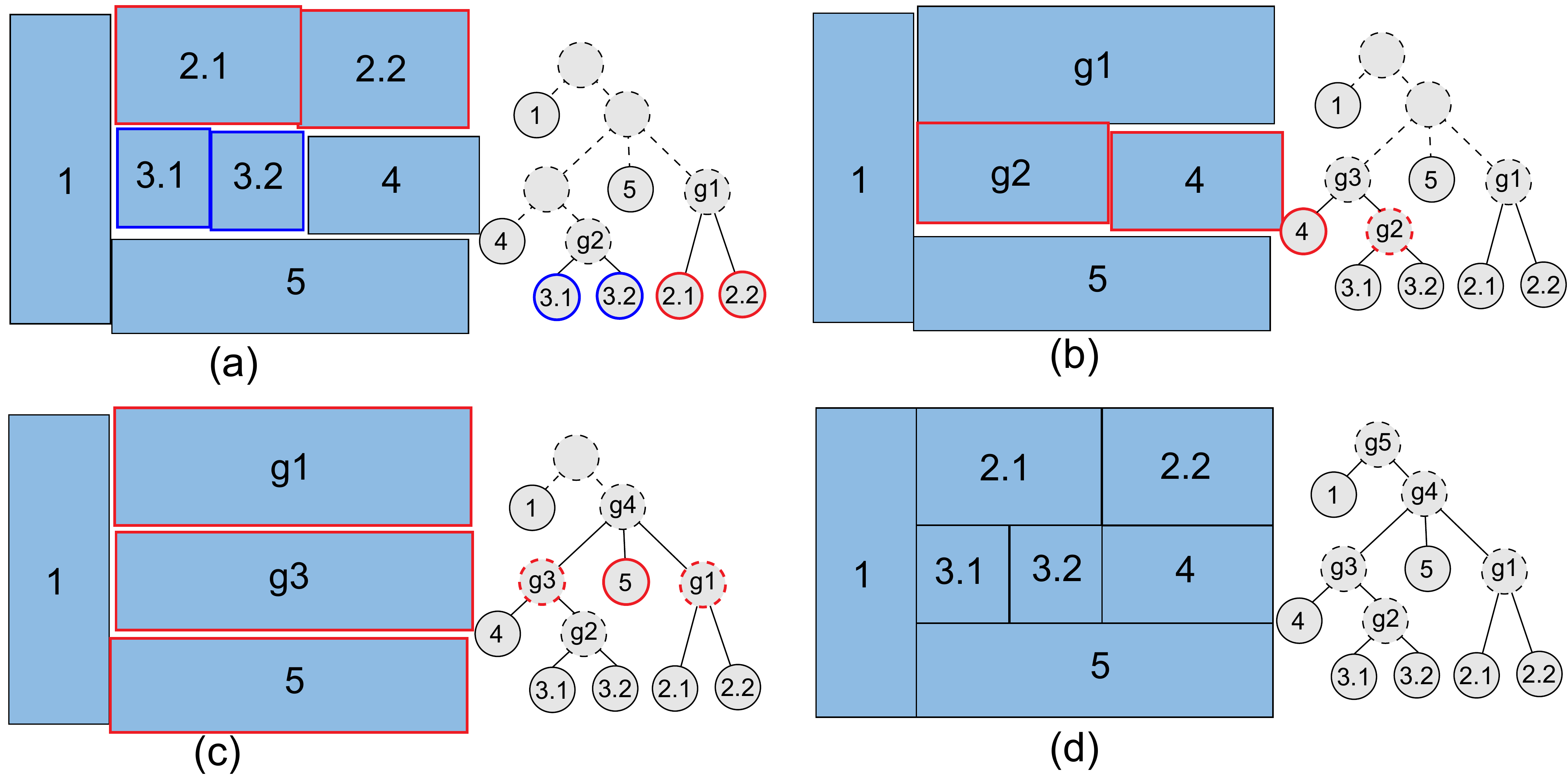}
  \vspace*{-7mm}
  \caption{A bottom-up approach is applied to refine manual layout annotations:
  views in a small multiples are grouped and aligned (a);
  neighboring views that can form a rectangle are grouped and aligned (b\&c).
  The process is repeated until no more views can be grouped, generating the refined layout (d) without overlaps and gaps.}
  \label{fig:align_process}
  \vspace*{-5mm}
\end{figure}

Manual annotation would inevitably cause overlaps or gaps between $bbox$ of neighboring views.
To remove the effect on follow-up analyses, we refine $bbox$ annotations using a bottom-up approach, as illustrated in Figure~\ref{fig:align_process}.
Here, the approach takes a set of rectangles $BB := \{bbox_1, [bbox_{2.1}, bbox_{2.2}], [bbox_{3.1}, bbox_{3.2}], bbox_4, bbox_5\}$ as input.
Notice here $BB$ is the annotation results for the MV displayed in Figure~\ref{fig:label_tool}.
The algorithm works as follows:

\begin{enumerate}
\vspace{-2mm}
\item
First, the algorithm checks if any two or more views are forming a small multiples, by referring to the view ID information stored in the JSON file.
For example, $bbox_{2.1}$ and $bbox_{2.2}$ in Figure~\ref{fig:align_process}(a) are bounding boxes of two views forming a small multiples.
Next, we group the $bbox$es together, forming a group of $bbox$es (denoted as $BB_{g}$).
We identify a minimum rectangle $bbox_g$ that encloses all $bbox \in BB_{g}$.
As in Figure~\ref{fig:align_process}, we can derive $bbox_{g1}$ for $BB_{g1} := \left\{bbox_{2.1}, bbox_{2.2}\right\}$, and $bbox_{g2}$ for $BB_{g2} := \left\{bbox_{3.1}, bbox_{3.2}\right\}$.

\vspace{-2mm}
\item
Next, we update $bbox_i \in BB_g$ as follows:
(i) in case if the top/left/bottom/right margin of $bbox_i$ to $bbox_g$ is smaller than a threshold $\theta$, we align $bbox_i$ to the top/left/bottom/right of $bbox_g$;
(ii) in case if an overlap or gap between two neighboring boxes $bbox_i$ \& $bbox_j$ is smaller than $\theta$, we stretch or shrink $bbox_i$ and $bbox_j$ to remove the misalignment.
Here, we set $\theta$ as 3\% of the average width and height of $bbox_g$.

\vspace{-2mm}
\item
All $bbox_i \in BB_g$ are removed from $BB$ while $bbox_g$ is added into $BB$, forming a new set $BB'$.
In Figure~\ref{fig:align_process}(b), $BB' := \{bbox_1, bbox_{g1}, bbox_{g2}, bbox_4, bbox_5\}$.

\vspace{-2mm}
\item
We then check if any two or more boxes in $BB'$ can be grouped upon the following conditions:
\begin{itemize}
\vspace{-2mm}
\item
The boxes are in the same \textit{``neighborhood''} $-$ centers of the boxes can be connected in a straight line without crossing some other box.
For instance, $bbox_{g2}$ \& $bbox_4$ are neighbors, while $bbox_{g1}$ \& $bbox_5$ are not because connections between them will pass through $bbox_{g2}$ and $bbox_4$.

\vspace{-0.5mm}
\item
The box centers are in horizontal or vertical, and widths/heights of the boxes are nearly the same.
For instance, $bbox_{g2}$ \& $bbox_4$ satisfy the condition, while $bbox_{g2}$ \& $bbox_1$ because their heights are very different.
\end{itemize}
\vspace{-2mm}
As in Figure~\ref{fig:align_process}(b), only $bbox_{g2}$ \& $bbox_4$ meet the conditions.

\vspace{-2mm}
\item
We update the boxes as described in Step 2.
Specifically, all sub-boxes in a group box, \emph{e.g.}, $bbox_{3.1}$ \& $bbox_{3.2}$ in $bbox_{g2}$ will be updated accordingly.

\vspace{-2mm}
\item
We repeat Step 2-5 until all boxes are grouped.
\end{enumerate}

\vspace{-2mm}
Finally, we generate a refined layout (Figure~\ref{fig:align_process}(d)) without overlaps and gaps for each MV.
We refine all interfaces in the annotation dataset.
Supplementary Table S3 presents some annotation results.

\section{Composition and Configuration Analysis}
\label{sec:analysis}

\begin{figure}[t]
  \centering
  \includegraphics[width=0.475\textwidth]{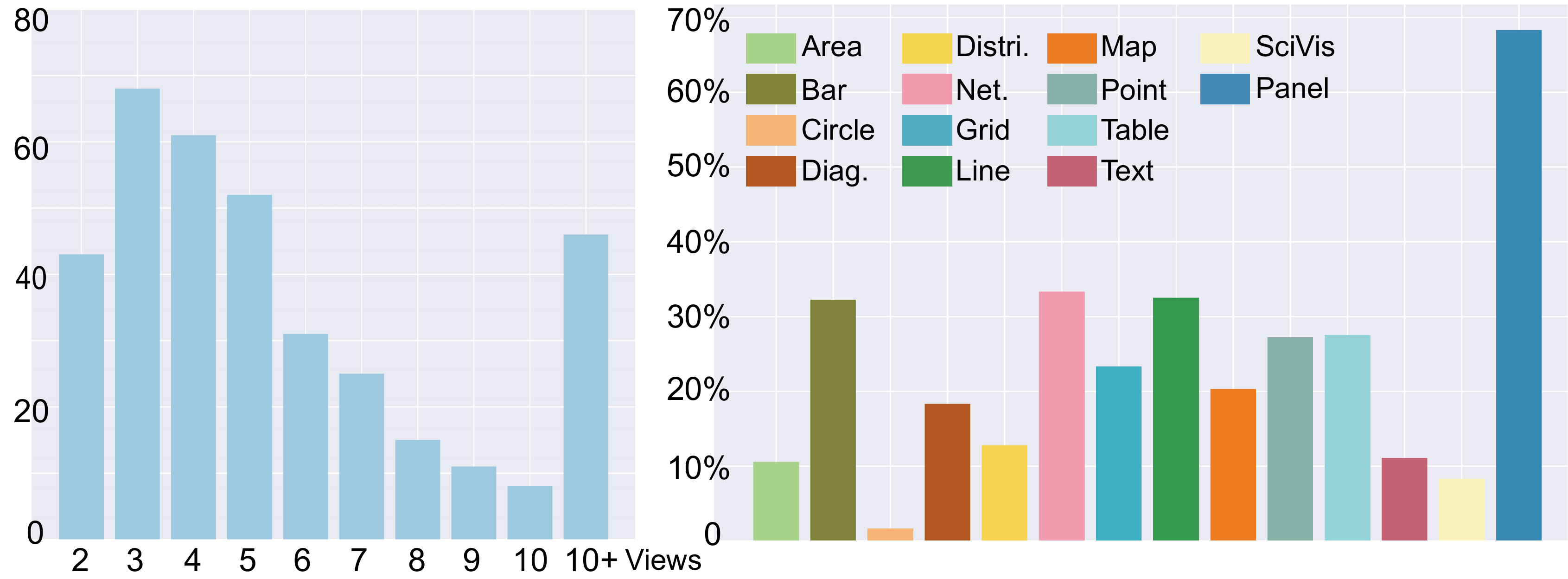} \\{}
  \vspace*{-4mm}
  \caption{Preliminary analyses on distribution of view count (left), and frequency of each view type (right). 
  }
  \label{fig:vis_stats}
  \vspace*{-5mm}
\end{figure}

After annotating all collected MV images (Section~\ref{sec:data}), we create a new corpus $C_{mv} = \left\{MV_i\right\}_{i=1}^n$, where $n = 360$ is the number of collected MVs in this work.
Each view $view_i$ is represented as a two-tuple $view_i := (type_i, bbox_i)$.
We first conduct some preliminary analyses on the distribution of view count and frequency of view type.
Figure~\ref{fig:vis_stats} presents the results.
As shown in Figure~\ref{fig:vis_stats} (left), most MVs (62.2\%) comprise no more than five views, with 68 three-view being the most common MVs, followed by 61 four-view, 52 five-view, and 43 two-view.
This indicates that designers opt for simple MVs with a small number of views.
Figure~\ref{fig:vis_stats} (right) shows that most MVs include a \emph{Panel} (68.3\%) of menus, legends, and narrative texts.
Next to \emph{Panel}, we see \emph{SciVis} is not frequent (8.3\%), as there are not many SciVis MVs in the collected dataset.
Among view types of information visualization, \emph{Bar} (32.2\%), \emph{Net.} (33.3\%), and \emph{Line} (32.5\%) exceed 30\%, whilst \emph{Circle} charts are seldom (only 1.6\%) adopted.

However, though interesting, the preliminary analyses do not provide answers for practical questions such as ``which view types are frequently used together?'', or ``where to position each view?''.
We conduct further analyses on the composition pattern of view types (Section~\ref{ssec:Composition_Analysis}), configuration patterns of view layouts (Section~\ref{ssec:Configuration_Analysis}), and integrated composition and configuration patterns (Section~\ref{ssec:Integrated_Analysis}).

\subsection{Composition Pattern Analysis}
\label{ssec:Composition_Analysis}

The composition pattern is defined by the view types and their usage frequency in MV designs.
Figure~\ref{fig:vis_stats} (right) shows the frequency of individual view types, but not how view types are used together.
We compute pairwise relationships between two view types ($type_{i}$ and $type_{j}$) using conditional probability:

\vspace{-2mm}
\begin{equation}
\begin{aligned}
P(type_{i}|type_{j}) = P(type_{i}\cap type_{j}) / P(type_{j})
\end{aligned}
\end{equation}
\vspace{-3mm}

$P(type_{i}|type_{j})$ ranges from [0, 1]: close to 0 values indicate $type_{i}$ rarely appears in $MVs$ including $type_{j}$, whilst close to 1 values indicate $type_{i}$ always appears in $MV$s including $type_{j}$.
Note that $type_i$ and $type_j$ could be the same, which indicates the probability that the same view type appears two or more times in one $MV$.

\begin{figure}[t]
  \centering
  \includegraphics[width=0.475\textwidth]{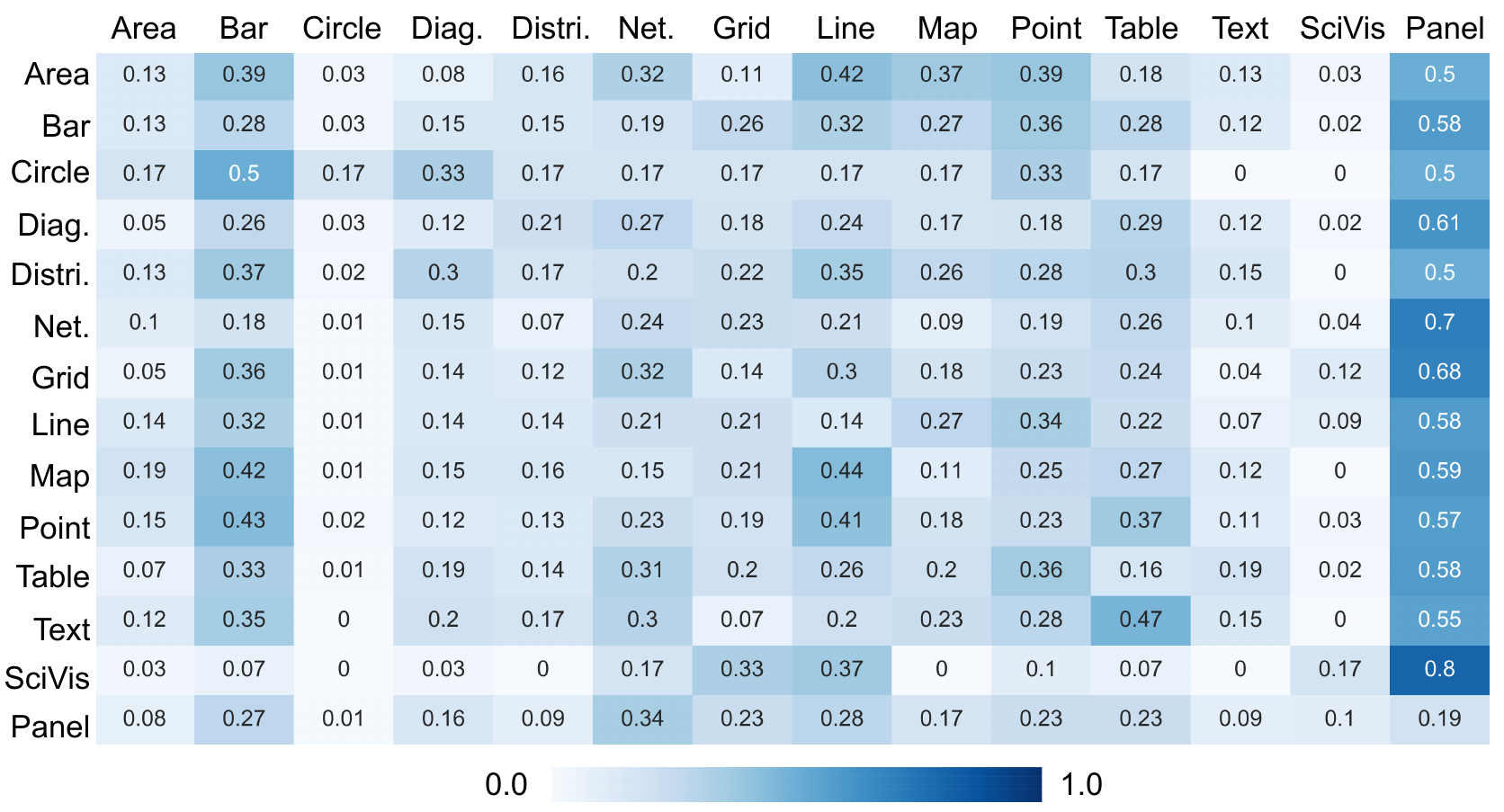} \\
  \vspace*{-4mm}
  \caption{
  Conditional probabilities of view types (columns) given other view types (rows) are employed in a MV.
  }
  \label{fig:simi_vt1}
  \vspace*{-5mm}
\end{figure}

Figure~\ref{fig:simi_vt1} presents the results, where $type_i$ is arranged in the columns and $type_j$ in the rows.
In the column of \emph{Panel}, the conditional probabilities are more than 0.5 given the other 13 view types, but $P$(\emph{Panel}|\emph{Panel}) is rather low, indicating it is rare to have two or more separate \emph{Panel}s in one $MV$.
We can also notice that the column of \emph{Bar} is quite dark, indicating that \emph{Bar} charts are frequently used together with other view types.
An exception is \emph{SciVis}, with $P$(\emph{Bar}|\emph{SciVis}) of 0.07, indicating that few \emph{SciVis} visualizations incorporate \emph{Bar} charts.
In contrast, \emph{SciVis} frequently adopts \emph{Panel}, as $P$(\emph{Panel} | \emph{SciVis}) = 0.8 is the highest value amongst all conditional probabilities.
Moreover, $P(type_{i}|type_{j})$ and $P(type_{j}|type_{i})$ can be very distinct.
For instance, $P$(\emph{Bar}|\emph{Circle}) reaches 0.5, whilst $P$(\emph{Circle}|\emph{Bar}) is only 0.03.
The difference is probably caused by the frequency differences between \emph{Circle} and \emph{Bar} charts (see Figure~\ref{fig:vis_stats} (right)).

\begin{figure}[t]
  \centering
  \includegraphics[width=0.475\textwidth]{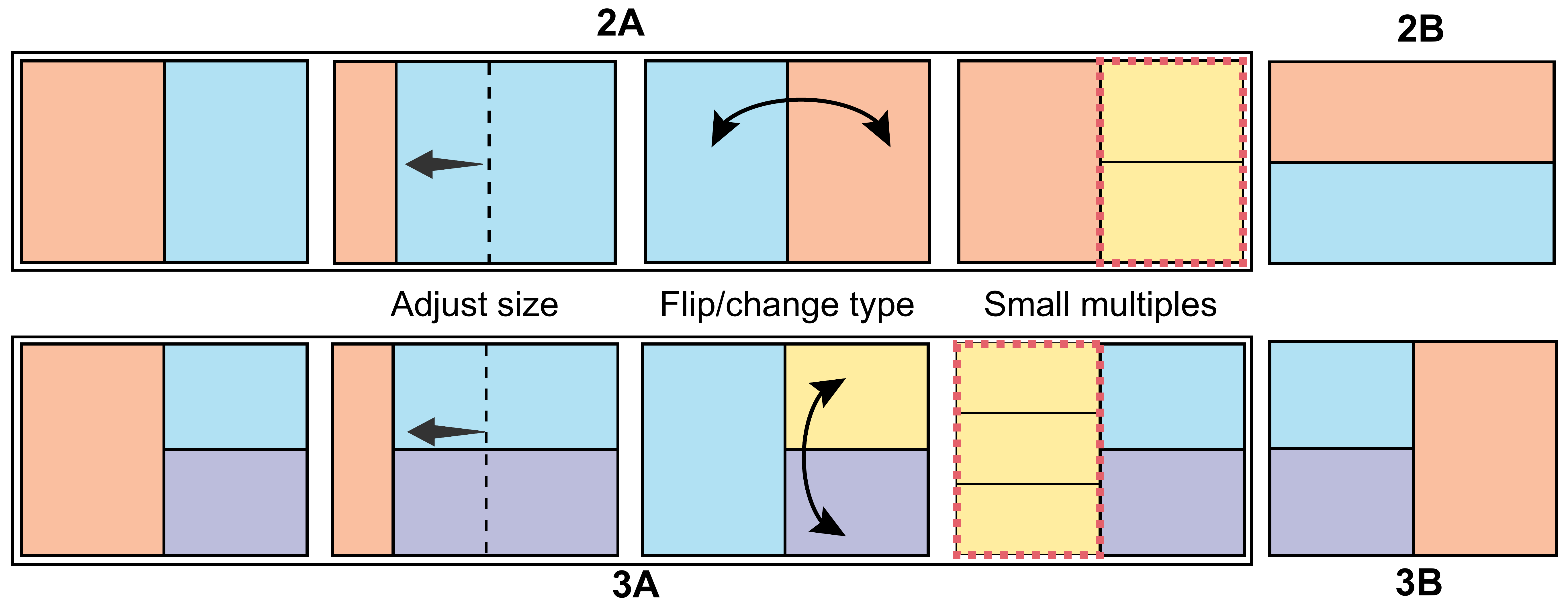} \\
  \vspace*{-3mm}
  \caption{
  Rules for coding MV layout:
  First, we assign a numeric number corresponding to the number of level-1 views/small multiples;
  Second, we assign an alphabetic character based on slice-and-dice order to distinguish layouts with the same view number.
  }
  \label{fig:code}
  \vspace*{-5mm}
\end{figure}

\subsection{Configuration Pattern Analysis}
\label{ssec:Configuration_Analysis}
The configuration pattern characterizes the spatial arrangement of view layouts in the display space.
We adopt a twofold coding rule to encode the configuration patterns of a MV design. 
In particular, each pattern is assigned two values, a numeric value followed by a letter (e.g. 2A, 3C, etc.). 
The rules of our coding scheme are described below.

\begin{itemize}
\vspace{-2mm}
\item
The numeric value of the coding rule refers to the number of ``top-level'' views in a MV (i.e. views that in the first level of the hierarchy). 
As in Figure~\ref{fig:code}, the MV designs in the top row will be assigned a 2, whereas the bottom ones will be assigned a 3.
 
\vspace{-2mm}
\item
The letter in the code refers to a specific type of layout.
Table~\ref{tab:stats} shows some of these examples. 
The letters themselves are not immediately meaningful. 
They are enumerated based on a slice-and-dice approach\cite{johnson_1991_tree-maps} to distinguish between different layouts found in our corpus.
For instance, the layout of \includegraphics[height=0.1in]{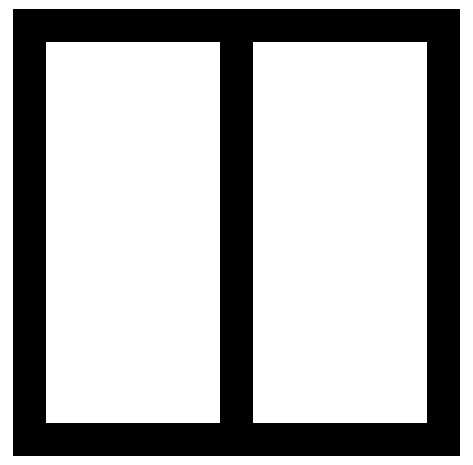} is accomplished by a vertical slice, while layout of \includegraphics[height=0.1in]{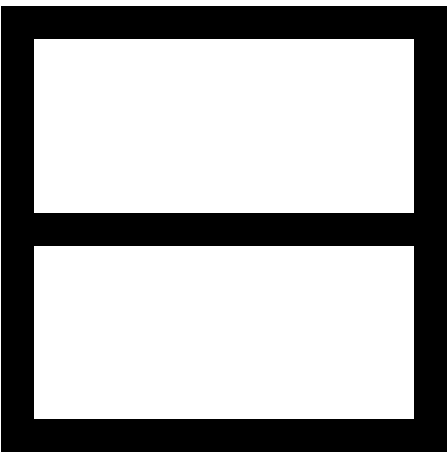} is accomplished by a horizontal slice. 
As such in Figure~\ref{fig:code}, the top one is assigned the letter `A' and the later the letter `B'.
Operations other than slicing, such as to adjust view size, and to flip or change view types, would not affect the slice-and-dice orders and are therefore disregarded.
When used with the numeric value, we produce the twofold encoding scheme (e.g., $2A$ for \includegraphics[height=0.1in]{figures/tile_img/1.png}, and $2B$ for \includegraphics[height=0.1in]{figures/tile_img/2.png}).
\end{itemize}

\begin{table}[t]
\caption{Top 10 layouts: numbers (green bars) and percentages (blue bar) in VAST, InfoVis, SciVis, EuroVis, and PacificVis.}
\begin{tabular}{c}
\begin{minipage}{0.485\textwidth}
\includegraphics[width=0.94\textwidth]{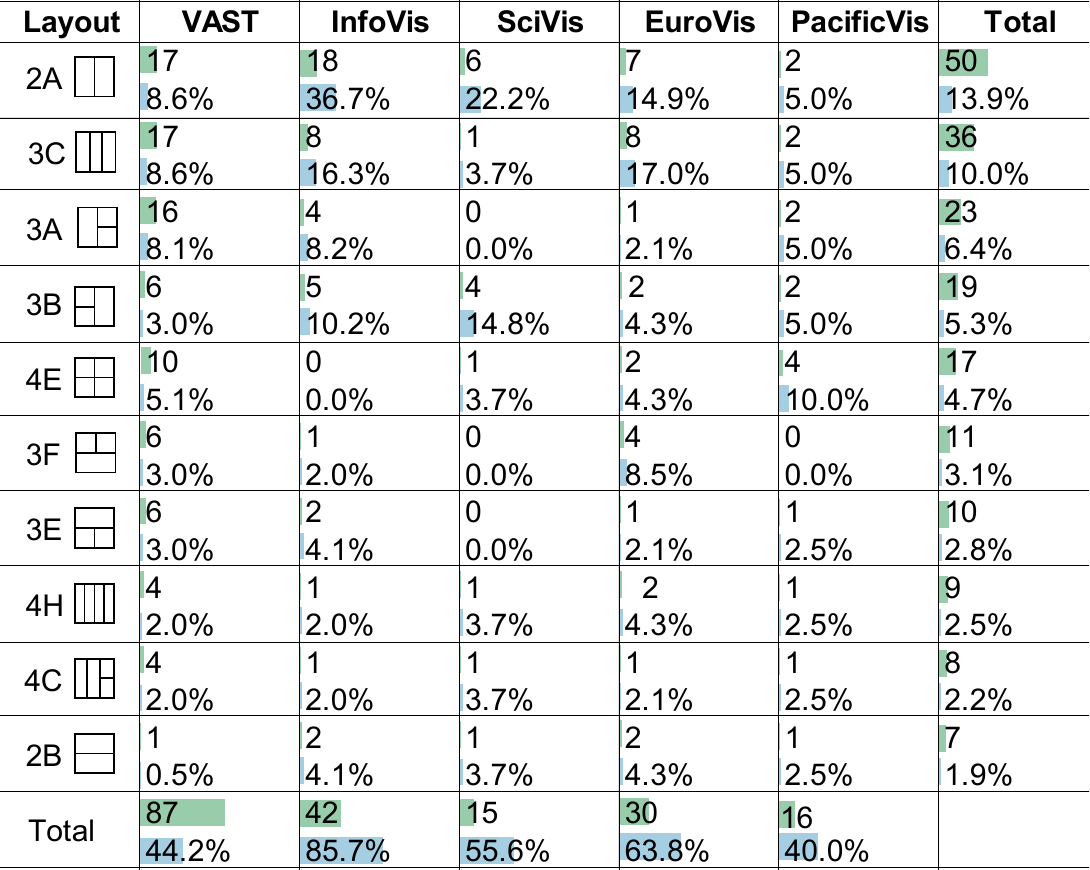}
\end{minipage}
\end{tabular}
\vspace{-5mm}
\label{tab:stats}
\end{table}

Based on this encoding scheme, we identify 98 unique layouts from 360 MV designs.
Table~\ref{tab:stats} presents the top 10 layouts, and the frequency of their uses (both in counts and percentages)
in VAST, InfoVis, SciVis, EuroVis, and PacificVis.
The result suggests that most MV designs adopt simple layouts, as the top 10 MV layouts all employ four or fewer views.
A closer investigation reveals differences between the conferences:
The top 10 layouts account for 85.7\% in InfoVis, followed by 63.8\% in EuroVis, 55.6\% in SciVis,  44.2\% in VAST, and least 40\% in PacificVis;
and $2A$ occupies higher percentages in InfoVis (36.7\%) and SciVis (22.2\%) than the other three conferences.
The findings suggest that InfoVis for abstract data, and SciVis for scientific data, tend to adopt simpler layouts, while VAST for data analysis requires more views to convey data from multiple perspectives.
In addition, even though EuroVis and PacificVis are both venues for all disciplines, EuroVis tends to show more preference for simpler layouts.

\subsection{Integrated Composition and Configuration Analysis}
\label{ssec:Integrated_Analysis}

We further conduct integrated composition and configuration analysis, aiming to reveal the relationship between view types and layouts.
Each view $view_i$ embraces two-tuple $(type_i,bbox_i)$, where $bbox_i$ consists of position $(x, y)$ and size $w \times h$.
From the data, we can associate view type with the position, and view type with size.
Note that absolute view sizes and positions are dependent on the display size, which is not accessible from MV images.
Instead, we analyze aspect ratios (Section~\ref{sssec:type_size}) and relative positions (Section~\ref{sssec:type_pos}) of view types.

\subsubsection{View Type \& Aspect Ratio}
\label{sssec:type_size}

\begin{figure}[t]
  \centering
  \includegraphics[width=0.475\textwidth]{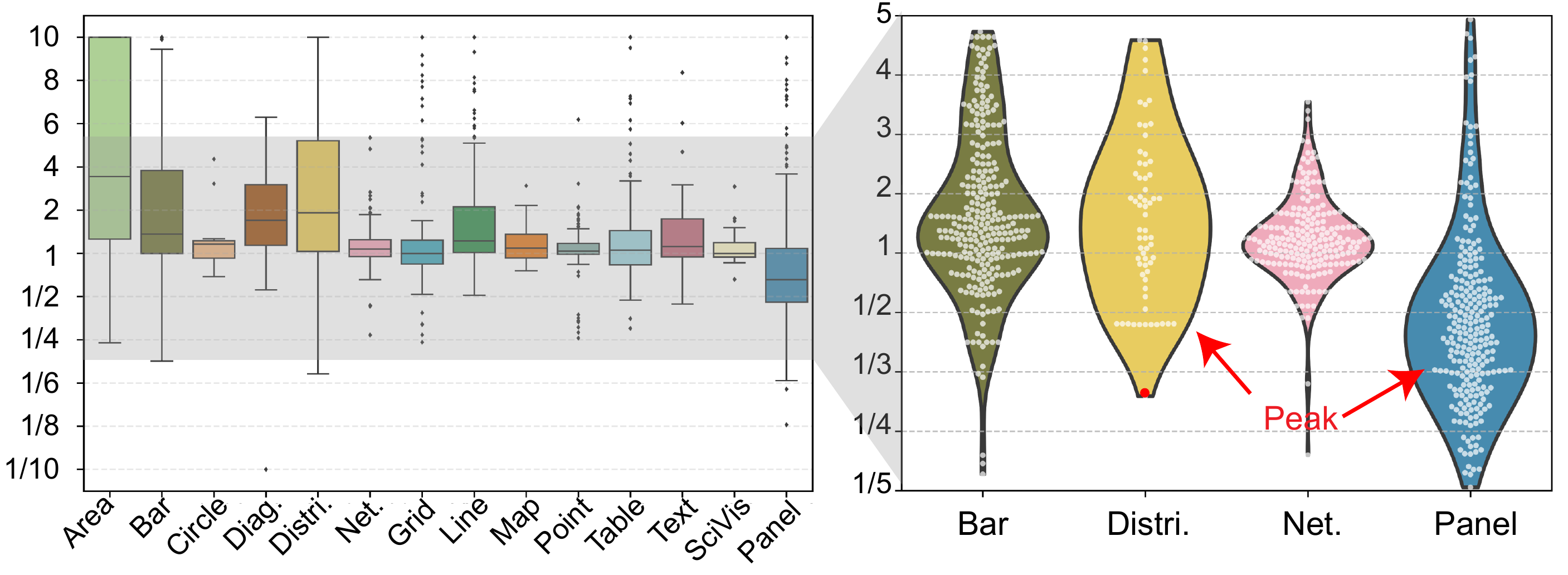} \\
  \vspace*{-4mm}
  \caption{
  Box plot (left) overviews aspect ratios of all 14 view types, and violin plot (right) depicts detailed aspect ratios of \emph{Bar},
  \emph{Distribution} (\emph{Distri.}), \emph{Tree and Network} (\emph{Net.}), and \emph{Panel} within the range [1/5, 5].
  }
  \label{fig:viewTypeRatio}
  \vspace*{-5mm}
\end{figure}

We first compute aspect ratio ($ARC$) of a view $view_i$ as $ARC_i = w_i/h_i$.
$ARC$ ranges from (0, +$\infty$): the value of 1 corresponds to a view in square, while values close to 0 or towards +$\infty$ indicate that the views are vertically or horizontally long and narrow, respectively.
Next, we group all views according to their view types, yielding 14 groups of $ARC$ values.
For each group, we depict its $ARC$ distribution using the box plot as in Figure~\ref{fig:viewTypeRatio} (left).
Here we show only the range [1/10, 10] that most $ARC$ values fall in.
Since $ARC$ and $1/ARC$ are reciprocal, we put 1 at the center, and [1/10, 1) and (1,10] as symmetric around 1.
We can observe that $ARC$s of most view types are within the range [1/2, 2], with mean values fall around 1.
Some exceptions are \emph{Area} charts with most $ARC$s larger than 1, and \emph{Panel} with most $ARC$s less than 1.
From follow-up investigation, we find that many MVs arrange horizontally long and narrow \emph{Area} charts in small multiples, and many MVs employ vertically long and narrow \emph{Panels} on the left/right side. 

We further select four representative view types of \emph{Bar}, \emph{Distri.}, \emph{Net.} and \emph{Panel}, and observe their $ARC$ distributions using the violin plot (Figure~\ref{fig:viewTypeRatio} (right)).
We can notice that the mean $ARC$ of \emph{Bar} charts is near 1, yet there are some narrow \emph{Bar} charts with $ARC$s towards 1/5 and 5.
For \emph{Distri.} views, there is a peak of $ARC$s (the white dots) around 1/2, and most $ARC$s are above 1/3 except one outlier. 
For \emph{Net.} views (e.g. node-link diagrams), we see their $ARC$s are more concentrated within [1/2, 2] with a mean value around 1.
This is probably because $ARC$s of \emph{Net.} views are more independent with the underlying data, in contrast to other view types like \emph{Bar} charts that need to increase its width to accommodate for the increased number of data attributes.
Lastly, we can see $ARC$s of \emph{Panel} are mostly below 1 with a peak around 1/3, which indicates that the \emph{Panel}s are typically vertically long and narrow.

\subsubsection{View Type \& Relative Position}
\label{sssec:type_pos}

\begin{wrapfigure}{R}{0.11\textwidth}
\raggedleft
\vspace{-5mm}
\includegraphics[width=0.11\textwidth]{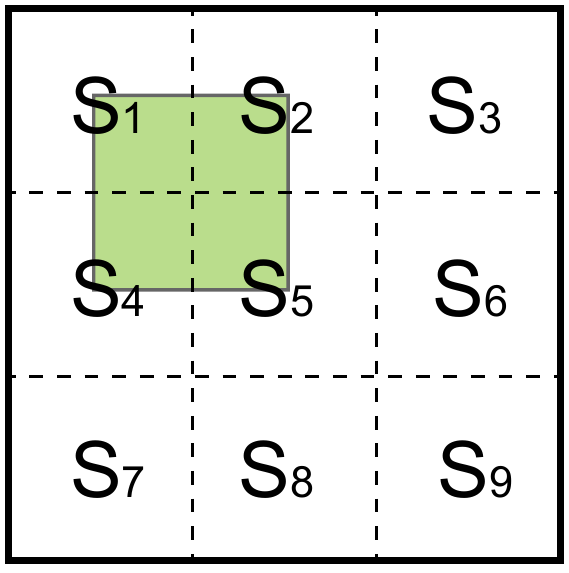} 
\label{fig:space_division}
\vspace{-6mm}
\end{wrapfigure}

\begin{figure*}[t]
  \centering
  \includegraphics[width=0.98\textwidth]{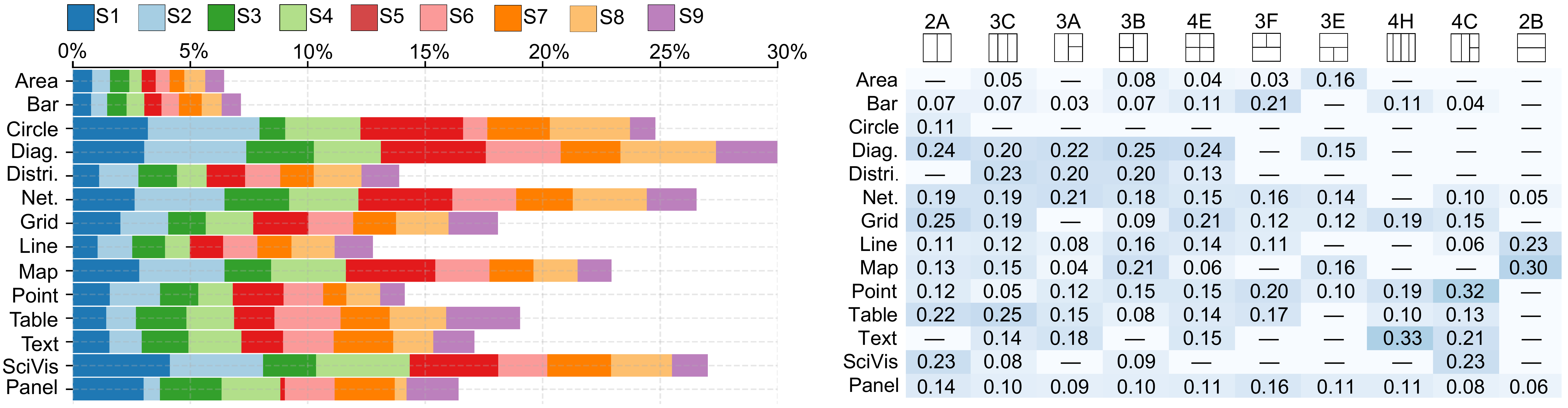} \\
  \vspace*{-4mm}
  \caption{
  Relative position of each view type averaged from all MVs (left), and
  stability of each view type in the top 10 layouts (right).
  }
  \label{fig:position}
  \vspace*{-4mm}
\end{figure*}

\emph{Position}.
To understand how individual views are positioned in MVs, we measure the relative positions of different view types.
Inspired by stability measurement for treemap layout\cite{sondag_2018_stable}, we divide the display space into $3\times3$ grids $\{S_1, S_2, ..., S_9 \}$ as shown in the inline figure.
Next, we model the relative position of $view_i$ in the display space $S$ as
$S(view_i) = \{p_{i,1} S_1, ..., p_{i,9} S_9 \}$, where $p_{i,k}$ stands for the proportion of overlapping area between $bbox_i$ and grid $S_k$ multiplied by the size of $S_k$.
For instance, consider the inline figure, the overlapping area between the green view (denoted as $view_1$) and $S_1$ is 1/4 the size of $S_1$, while $S_1$ is 1/9 the size of entire view space $S$.
Hence, $p_{1,1} = 1/4 \times 1/9 = 1/36$.
Similarly, we can compute $p_{1,2}$, $p_{1,4}$, and $p_{1,5}$ as 1/36, while the other $p_{1,j}$ are 0.
Together, we can represent the relative position of the green view as $\{\frac{1}{36}S_1, \frac{1}{36}S_2, \frac{1}{36}S_4, \frac{1}{36}S_5\}$.

The use of this encoding scheme has several advantages:
first, each view is explicitly represented $-$ we can derive relative positions and sizes of different views in a MV.
Second, the representation is consistent across MVs $-$ we can compare positions and sizes of views in different MVs. 
Last, we can sum up multiple views of the same view type by simply adding up their $p$ values.
The sum stands for relative position and size for the view type rather than a specific view.

The stacked bar chart in Figure~\ref{fig:position} (left) depicts the average $p$ values for each view type in MV designs.
The overall bar length depicts the average size of each view type.
We can notice that several view types, including \emph{Diag.}, \emph{SciVis}, and \emph{Net.}, occupy larger areas $-$ over 25\% of the display space.
In contrast, \emph{Area} and \emph{Bar} occupy only $\sim$5\% of the display space, even though the frequency of \emph{Bar} charts are quite high (see Figure~\ref{fig:vis_stats} (right)).
This may suggest that designers tend to assign small spaces for \emph{Area} and \emph{Bar} charts, as the view types are typically used for depicting summary statistics. 
We expected \emph{Circle} to show similar $p$ values with those of \emph{Area} and \emph{Bar}, but surprisingly \emph{Circle} occupies much bigger space of about $25$\%.
We investigated carefully the MVs containing \emph{Circle} in the database and found that works on infographics design use \emph{Circle} charts to illustrate their approach; see Supplementary Table S3-2A for an example.
In contrast, \emph{Circle} occupies much smaller sizes in other visual analytics MVs; see Supplementary Table S3-Other Layouts for an example.

From individual bars, we can derive relative positions of each view type.
We notice that \emph{Panel} is rarely distributed in the center of the display space, as its $p$ values for grids $S_2$, $S_5$, and $S_8$ are very small.
In contrast, \emph{Diag.} is mostly placed in the center, as its $p$ values for grids $S_2$, $S_5$ and $S_8$ are relatively high.
In addition to \emph{Diag.}, the other large view types, \emph{i.e.}, \emph{SciVis} and \emph{Map}, also show higher $p$ values at grids $S_1$, $S_2$, $S_4$, and $S_5$.
This indicates that designers tend to place large views in the top-left and center regions of the display space.
The other view types present balanced $p$ values for the nine grids, indicating their positions can be inconsistent depending on designs.

\vspace{1mm}
\noindent
\emph{Position Stability}:
To check if the relative positions of a view type can, in fact, be inconsistent, we measure position stability of a view type in different MVs.
We adopt the metric of \emph{relative-position-change}\cite{sondag_2018_stable} to measure the distance between two views $view_i$ and $view_j$:
\vspace{-2mm}
\begin{equation}
D(view_i, view_j) = \frac{1}{2}\sum_{k=1}^9|p_{i,k} - p_{j,k}|
\end{equation}
\vspace{-2mm}

$D$ ranges from [0, 0.5]: close to 0 values indicate similar, consistent positions across MV designs, whilst close to 0.5 values indicate positions are exclusively distinct (i.e. highly inconsistent).
From the \emph{relative-position-change}, we can derive stability (denoted as $STB$) of $type_k$ in the collected $MV$s as follows:

\vspace{-2mm}
\begin{equation}
STB(type) = \frac{1}{m(m-1)}\sum_{i=1}^m \, \sum_{j=1 , j \neq i}^m D(view_i, view_j)
\end{equation}
\vspace{-2mm}

\noindent
where $m$ indicates the number of $MV$s that include the view $type$.

Figure~\ref{fig:position} (right) presents the stability of each view type in the top 10 layouts.
The cells with null values indicate that the layout contains at most one $MV$ with the view type.
We can observe that most \emph{STBs} are below 0.2.
For example, \emph{STB}s of \emph{Panel} are about 0.1 in all top 10 layouts, indicating positions and sizes of \emph{Panel} are rather stable.
This is probably because designers commonly allocate \emph{Panel} in the periphery.
Nevertheless, there are several unstable cases with \emph{STB}s over 0.3, including \emph{Text} in layout \emph{4H}, \emph{Point} in layout \emph{4C}, and \emph{Map} in layout \emph{2B}.
Taking \emph{Map} in layout \emph{2B} for example, we notice that chances of positioning maps in the top or the bottom are almost half and half.
In contrast, \emph{Map} in layout \emph{2A} is rather stable, as we see most maps are positioned in the left, rather than in the right.
\section{Recommendation System}
\label{sec:app}

Our analyses are nuanced and therefore difficult to turn into a cleanly articulated design guideline for MVs. 
As such, we aggregate our findings into a recommendation system that can help a designer choose the most appropriate MV designs given their data and needs.
Our recommendation system can be used in two interactive modes: \emph{Exploration} mode (Section~\ref{ssec:exp_mode}) and \emph{Design} mode (Section~\ref{ssec:design_mode}).

\subsection{Exploration Mode}
\label{ssec:exp_mode}
\emph{Exploration} mode enables faceted exploration of existing MVs.
Based on the prior analyses, each MV includes attributes of \emph{view types}, \emph{number of views}, and \emph{layout}.
Moreover, the MV images are derived from publications that include attributes of \emph{year}, \emph{venue}, \emph{authors}, etc.
We develop \emph{Exploration} mode as illustrated in Figure~\ref{fig:exp_mode}, which consists of two main components:

\begin{itemize}
\item
\textbf{Exploration View}, shown in Figure~\ref{fig:exp_mode}(b), adopts a unit visualization to present the query result.
Each existing MV design is depicted as a dot, with color representing extrinsic attributes of \emph{year} or \emph{venue}.
Users can group the queried MVs based on their intrinsic attributes of \textit{Number of Views} or \textit{Layout}.
We only present the first 10 groups in case more than 10 groups are formed.
Here we omit \emph{View Type} since a MV typically consists of multiple view types.
Coloring and grouping options are provided as a drop-down list.

Details of a $MV$ is shown when the mouse hovers over the dot; see Figure~\ref{fig:exp_mode}(b1) for an example. 
The detail view consists of two perspective information:
first, the $MV$ image is presented on the left side;
second, the publication metadata, including \emph{title}, \emph{authors}, \emph{doi}, and \emph{venue}, are presented on the right side.
Users can click on \emph{doi} that links to the publication website.

\item
\textbf{Query Panel}, shown in Figure~\ref{fig:exp_mode}(a), consists of three faceted panels $-$ \emph{View Type} (T), \textit{Number of Views} (N), and \textit{Layout} (L), which are the essential elements of composition and configuration patterns.
Each panel comprises of all possible attribute values, \emph{e.g.}, \emph{Bar} \& \emph{Line} in \emph{View Type}, \emph{2 to 10+} in \emph{Number of Views}, and \includegraphics[height=0.1in]{figures/tile_img/1.png} \& \includegraphics[height=0.1in]{figures/tile_img/2.png} in \textit{Layout}.

Using the query panel, a user can filter and select the MV designs that meet their goals.
Users can select multiple attribute values:
in case the attribute values fall in the same panel, we take the union of filtering results by individual attribute value;
in case the attribute values fall in different panels, we take the intersection of filtering results by individual attribute values.

\begin{figure}[t]
  \centering
  \includegraphics[width=0.495\textwidth]{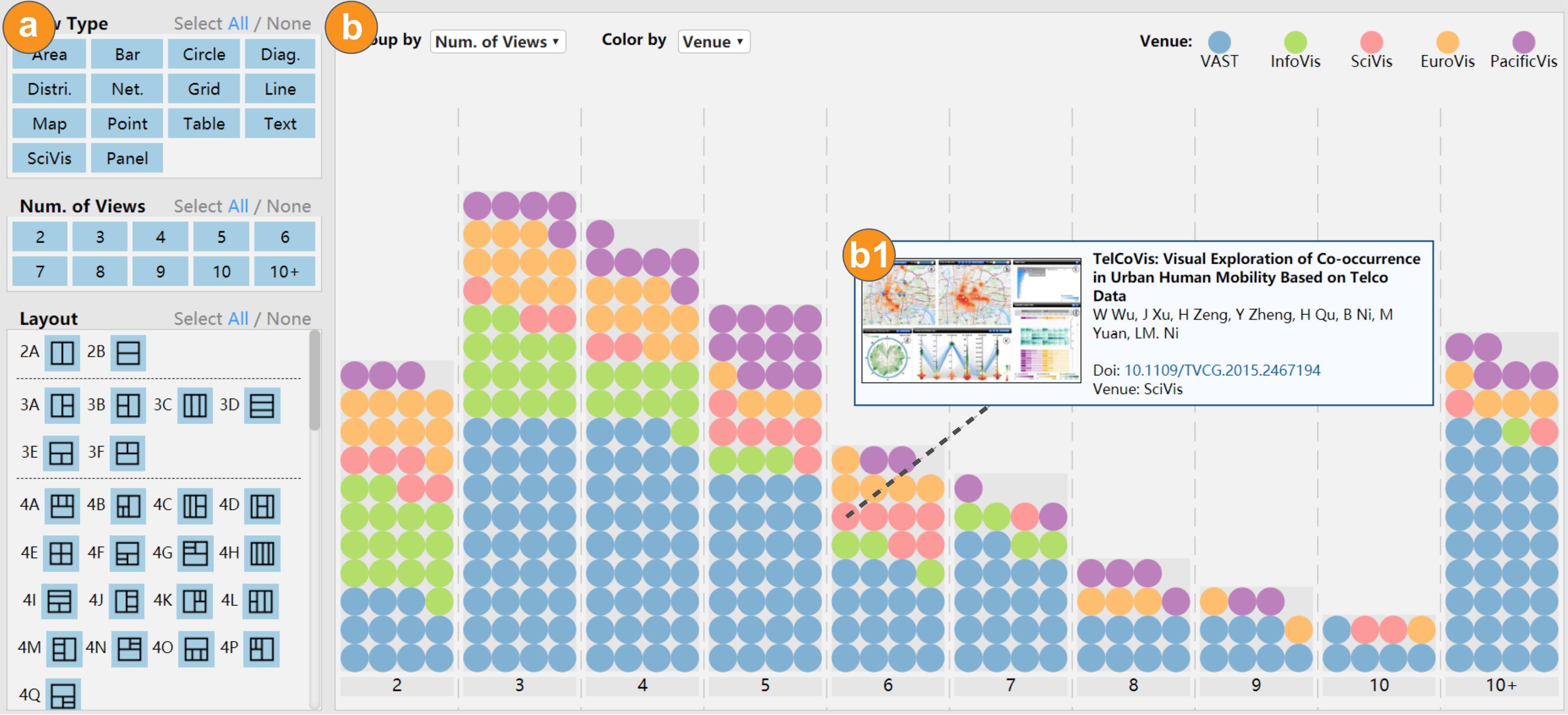} \\
  \vspace*{-3.5mm}
  \caption{
  Exploration mode mainly consists of \emph{Query Panel} (a) that enables multi-faceted query, and \emph{Exploration View} (b) that depicts the query results using unit-based visualization.
  }
  \label{fig:exp_mode}
  \vspace*{-5mm}
\end{figure}

\begin{figure*}[t]
  \centering
  \includegraphics[width=0.95\textwidth]{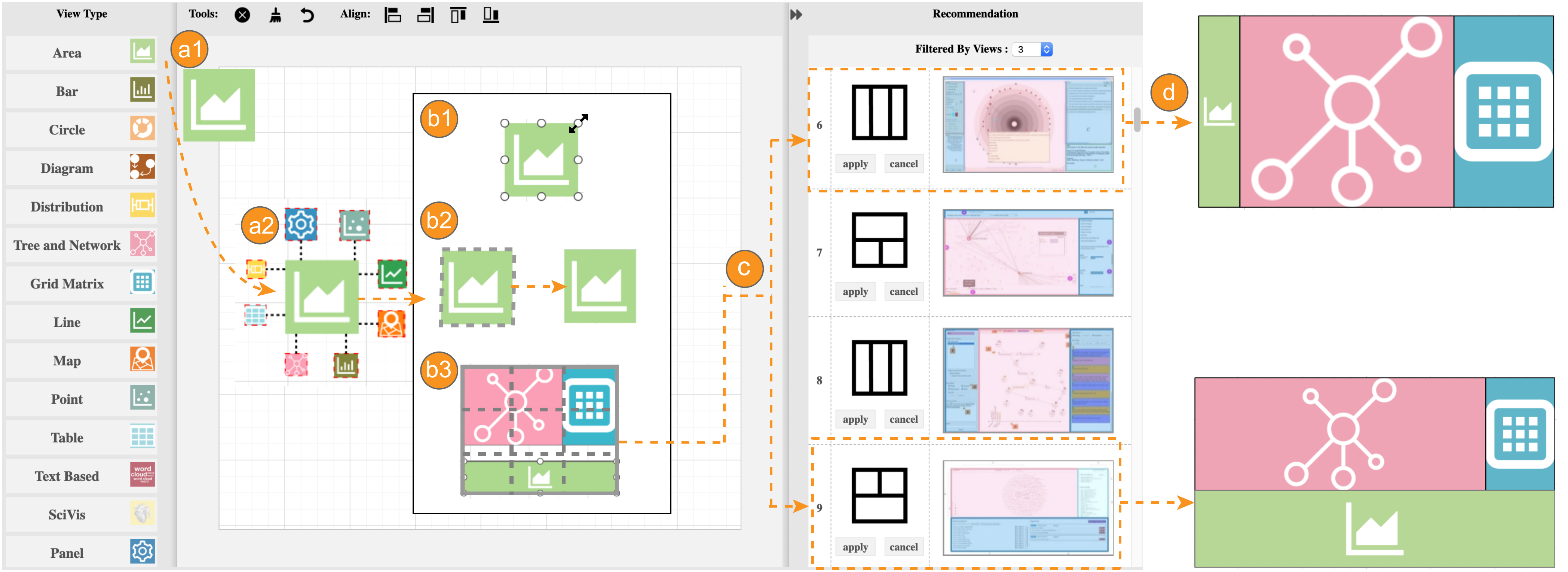} \\
  \vspace*{-3.5mm}
  \caption{
  Design mode incorporates a set of operations to support interactive design: to add view through drag-and-drop (a1), or icon selection (a2); to adjust view size (b1) or position (b2), or align multiple views together (b3); to recommend design based on configuration and composition proximity with existing designs (c).
  A recommended design can be further refined based on user preference (d).
  }
  \label{fig:design_mode}
  \vspace*{-5mm}
\end{figure*}

\end{itemize}

Figure~\ref{fig:exp_mode} presents a screenshot of our recommendation system in \emph{Exploration} mode.
Here, the query result is all MVs in the dataset, since all attribute values are selected.
The grouping option is \emph{Number of Views}, and the same distribution with that in Figure~\ref{fig:vis_stats} (left) is presented.
The coloring option is \emph{Venue}: the most of dots are colored as light blue (\emph{VAST}), and amounts of dots in the other four colors (\emph{InfoVis, SciVis, EuroVis}, and \emph{PacificVis}) are close. 
In Figure~\ref{fig:exp_mode}(b1), a dot corresponding to the MV in\cite{Wu_2016_TelCoVis} is highlighted, which is a SciVis paper utilizing six views to present urban informatics.

\subsection{Design Mode}
\label{ssec:design_mode}

Our recommendation system can also be used in the \emph{Design} mode to recommend $MV$ designs.
The idea behind the design mode is for a user to ``draw'' their desired MV design, and the system will search through the database of MV designs to recommend the closest existing designs. Under the hood, the recommendation system takes into account the position of the individual views, their sizes, and the overall composition in finding the best matches.
As shown in Figure~\ref{fig:design_mode}, the recommendation system (in the Design Mode) consists of three views: \emph{View Type List}, \emph{Design Panel}, and \emph{Recommendation Gallery}.
The system supports the following operations:

\begin{itemize}
\vspace{-2mm}
\item
\textbf{Add/remove view}:
Users can add a view by dragging-and-dropping a view icon from \emph{View Type List} to \emph{Design Panel} (Figure~\ref{fig:design_mode}(a1)).
A selected view will be surrounded by multiple view icons, with icon size corresponding to their correlations to the view of selection.
Users can add a second view that is closely related to the selected view by clicking on the surrounding icon.
Added views can be removed by clicking on the \emph{delete} or \emph{remove all} button in the top.

\vspace{-2mm}
\item
\textbf{Adjust view}:
Users can adjust the size of a selected view by dragging handles at its sides and corners (Figure~\ref{fig:design_mode}(b1)) and adjust its position by dragging (Figure~\ref{fig:design_mode}(b2)).
Further, users can perform the ``alignment'' operation\cite{xu_2014_global} to align several selected views  (Figure~\ref{fig:design_mode}(b3)).

\end{itemize}

\vspace{-2mm}
With the above operations, users can already design the layout of a MV.
However, the generated layouts may not reflect the good practice of MV designs.
To suggest good and similar MV designs, our system searches through the database with the user-designed layout as input.
This kind of exemplar-based recommendation has been widely applied in designing infographics (\emph{e.g.},\cite{donovan_2014_learning, chen_2019_towards}), and mobile apps (\emph{e.g.},\cite{swearngin_2018_rewire}).
In our system, the recommendation algorithm works as follows:

\begin{itemize}
\vspace{-2mm}
\item

\textbf{Recommend layout}:
As described in Section~\ref{sssec:type_pos}, we divide the display space of a $MV$ into $3\times3$ grids.
Each grid comprises up to 14 view types, with each view type occupies a certain proportion of the grid.
By this, we can represent a $MV$ as a three-dimensional tensor $T_{mv} \in \mathbb{R}^{3 \times 3 \times14}$.
Similarly, we can regard user input of multiple views as a 3D tensor $T_{in} \in \mathbb{R}^{3 \times 3 \times14}$, by treating the minimum bounding box enclosing all views as the display space and dividing the space into 3$\times$3 grids (see Figure~\ref{fig:design_mode}(b3)).

Next, we employ mutual information ($MI$) as a quantitative indicator for measuring proximity between $T_{mv}$ and $T_{in}$.
Since $T_{mv}$ and $T_{in}$ are both divided into 3$\times$3 grids, they are geometrically aligned.
Hence, the measurement is simplified without geometry matching.
Before calculating the mutual information, we reshaped the $T_{mv}$ and $T_{in}$ to one-dimension vector, i.e., $1 * n$ and $n$ is 126.
Then we discretized the continuous data using Equal-Width discretization.
$MI$ is computed as

\vspace{-2mm}
\begin{equation}
MI(T_{mv}, T_{in}) = \sum_{i \in T_{mv}} \sum_{j \in T_{in}} P(i, j) log\frac{P(i,j)}{P(i)P(j)}
\end{equation}

\vspace{-2mm}
where $P(i)$ and $P(j)$ are the marginal probability distribution functions $-$ computed via normalized intensity
histograms $-$ of $T_{mv}$ and $T_{in}$ respectively, and $P(i, j)$  is the joint probability function of $T_{mv}$ and $T_{in}$.

We iterate over all 360 MVs in the database, compute their $MI$s to user inputs, and sort them in descending order of $MI$.
The sorted results are listed in \emph{Recommendation Gallery}, with layouts and MV images arranged side-by-side.
Users can filter the ordered MVs based on constraints of \emph{Number of Views}.
Layout of an example $MV$ can then be applied to user inputs of views.
Figure~\ref{fig:design_mode}(d) presents two example layouts applied to user inputs.

\end{itemize}

Note that we can divide a $MV$ into finer grids, \emph{e.g.}, $32\times32$, and represent a $MV$ as a 3D tensor $\mathbb{R}^{32\times32\times14}$.
The refinement can give better proximity between two MVs.
However it will increase the computation time, and the effects are marginal as we observed from experiment results, probably because the number of views in a $MV$ is limited.
Therefore we empirically select $3\times3$ as the granularity of the grid in our system.

\subsection{User Study}
\label{ssec:eval}

We conducted two studies to evaluate \emph{Exploration} mode and \emph{Design} mode of the recommendation system.
Due to the COVID-19 pandemic, both studies were performed virtually.

\subsubsection{Study 1: Qualitative Evaluation for Exploration Mode}

\textbf{Participants}: 
We recruited 20 participants (6 females, 14 males, age: 23.65$\pm$1.18) in the study.
The participants all held a Bachelor's degree in fields like computer science, finance, and engineering.
They were familiar with WIMP (Windows, Icons, Menus, and a Pointing device) interfaces, but none of them has designed a MV visualization before.

\noindent
\textbf{Procedure}: 
We first explained to the participants the domain-specific terms, e.g., MVs and view types.
Then we introduced functionalities of the \emph{Exploration} mode and gave the participants about 15 minutes to freely explore the interface.
After they felt comfortable using the tool, we asked the participants to find answers for 10 questions regarding composition and configuration patterns of MVs, \emph{e.g.}, \emph{``which layout is the most frequently used in MVs?''}, and \emph{``how many MVs of six views contain bar chart?''}.
The participants were asked to submit answers immediately when they felt confident with their solution.
After the main study, the participants were asked to finish a questionnaire (see Supplementary Table S4).
The questionnaire comprised three 7-point Likert scale questions regarding general impression (Q1), usefulness (Q2), and simplicity (Q3), and two free-form questions on what can be improved (Q4) and general feedback (Q5).
Answers for the 7-point questions range from strongly disagree (1) to strongly agree (7).

\begin{figure}[t]
  \centering
  \includegraphics[width=0.475\textwidth]{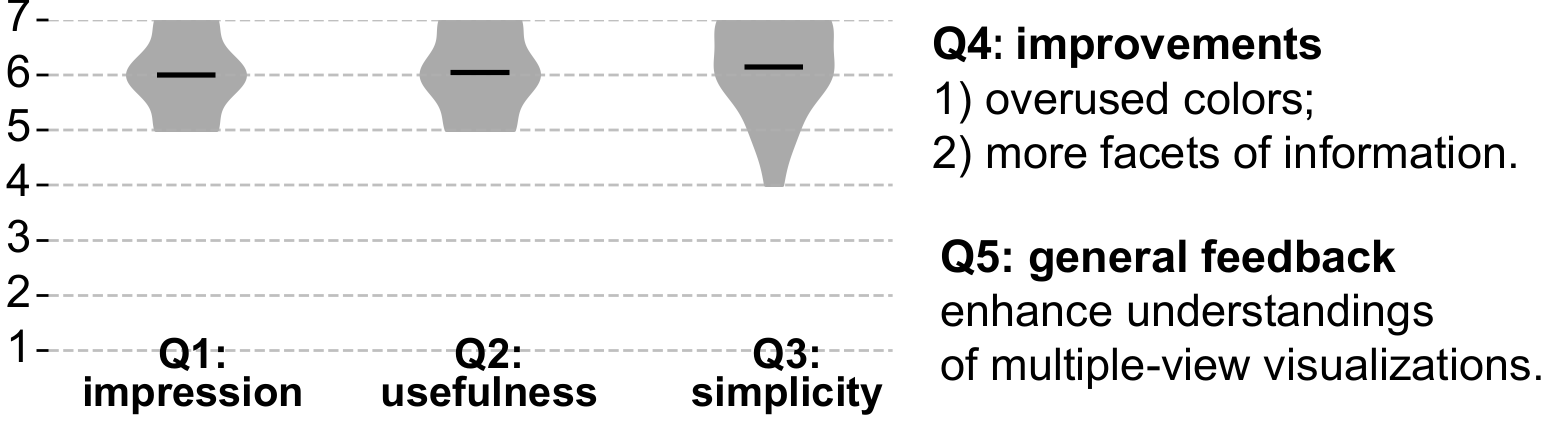} \\
  \vspace*{-4mm}
  \caption{
  Qualitative feedbacks for the \emph{Exploration} mode.
  }
  \label{fig:feedback}
  \vspace*{-5mm}
\end{figure}
\noindent
\textbf{Results}: 
All participants finished the 10 questions in eight minutes.
The accuracy of answers for each question is over 90\%, and overall the average accuracy is up to 97.8\%. 
Figure~\ref{fig:feedback} presents the results for the questionnaire.
Overall, the participants had an overall positive impression of the system (Q1) (mean = 6, SD = 0.75), found the system to be useful (Q2) (mean = 6.11, SD = 0.74), and perceived the \emph{Exploration} mode as easy to use (Q3) (mean = 6.21, SD = 0.85).
In the free-form question on improvements (Q4), two suggestions for improvement were raised:
(i) some colors are overused, and (ii) more facets of information can be integrated, such as \emph{authors} and \emph{research topics}.
The second aspect could be addressed in the near future by building a more comprehensive database of MV designs.

\subsubsection{Study 2: Quantitative Evaluation for Design Mode}

\textbf{Participants}:
We recruited 13 participants (2 female and 11 male) who are graduating Ph.D. students, PostDocs, and junior professors in data visualization.
All participants have experience in visualization design for at least four years.

\noindent\textbf{Procedure}: 
We first introduced the motivation of the work and functionalities of the \emph{Design} mode. 
Next, we gave the participants about 15 minutes to freely explore the interface.
After that, the participants were asked to complete three tasks of arranging 3 (Task 1), 5 (Task 2), and 7 (Task 3) views into a MV.
Three design modes were provided:
\begin{itemize}
\vspace{-2mm}
\item
\textit{Basic} mode only includes basic functions of \emph{add/remove view}, \emph{adjust view} by resizing and dragging;

\vspace{-2mm}
\item
\textit{Partial} mode includes an additional function of \emph{adjust view} by alignment; and

\vspace{-2mm}
\item
\textit{Full} mode further includes the function of \emph{recommend layout}.
\end{itemize}

\vspace{-2mm}
The completion time of each task and design mode was recorded.
The order of the design mode was randomly assigned to each participant in order to counter-balance learning effects.
In the end, participants were asked to fill out user feedback (Supplementary Table S5) on the general impression of the \emph{Design} mode. 

\noindent
\textbf{Hypotheses}:
We anticipate completion time is dependent on both the number of views and design mode. 
We formulated two hypotheses:
first (\textbf{H1}), the task of arranging 7 views will be slower than arranging 5 views, and arranging 5 views will be slower than arranging 3 views.  
We expect an increase in completion time when the number of views increases.
Second (\textbf{H2}), \textit{Full} mode is the fastest, followed by \textit{Partial} mode and \textit{Basic} mode. 
We expect alignment and recommendation functions will save time when designing MVs.

\begin{figure}[t]
  \centering
  \includegraphics[width=0.495\textwidth]{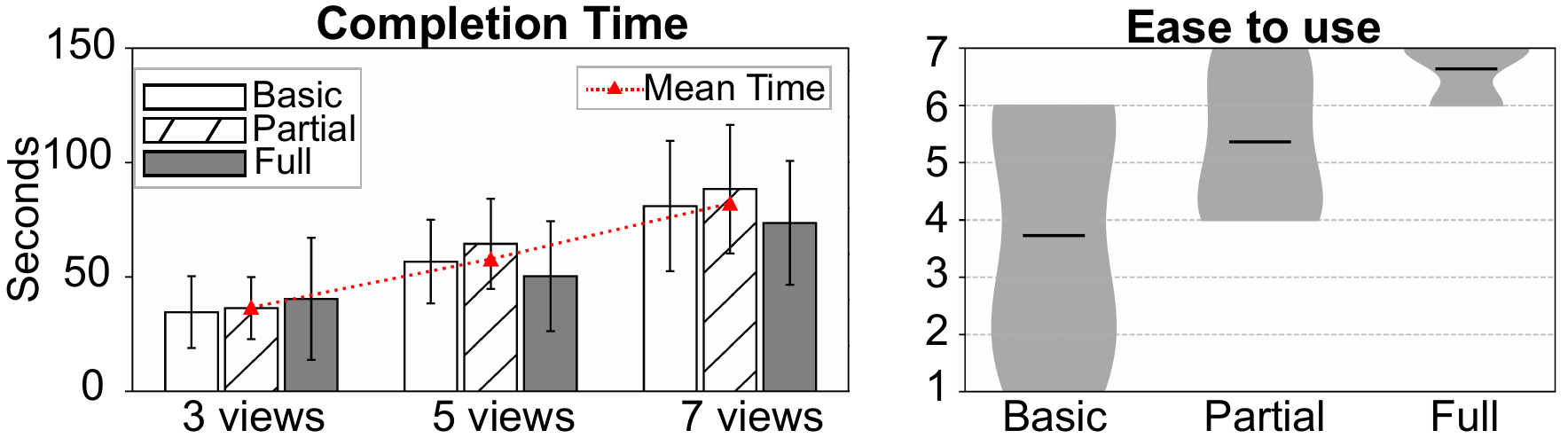} \\
  \vspace*{-4mm}
  \caption{
  Results of quantitative evaluation for \emph{Design} mode: completion time (left), and feedback on the easy of use (right).
  }
  \label{fig:design_rank}
  \vspace*{-5mm}
\end{figure}

\noindent
\textbf{Result}: 
We collected in total 3 (tasks) $\times$ 3 (modes) $\times$ 13 (participants) = 117 trails.
We removed abnormal trails of completion times exceeding two times than the others by two participants.
The remaining data were in line with the normal distribution after testing with Shapiro-Wilk test, 
and we performed a two-way ANOVA (3 tasks $\times$ 3 modes) to analyze the data.
Figure~\ref{fig:design_rank} (left) summarizes the result.
Significant effects of task complexity on completion time $(F(2,90) = 32.21, p < 0.001)$ were observed.
We further performed post-hoc comparisons of completion time among the number of views.
Task 1 is on average $36.7s$ faster than Task 2 $(p < 0.001)$, while Task 2 is on average $57.8s$ faster than Task 3 $(p < 0.001)$.
The results confirmed \textbf{H1}.
However, no significant effects of design mode on the completion time $(F(2,90) = 0.86, p > 0.1)$ were observed.
This result leads us to reject \textbf{H2}.

\noindent
\textbf{Observations}:
Though \textbf{H2} was not supported, we observed interesting design practices from the experiments.
We noticed that in \emph{Full} mode, some participants looked through recommended layouts, and tried out different layouts attempting to find an optimal one, whilst in \emph{Basic} and \emph{Partial} modes, the participants directly arranged the views by dragging and resizing.
Looking up and trying out example layouts in \emph{Full} mode cost additional time in completing the task, which explains why completion times of \emph{Full} mode are more diverse, especially for Task 1.
We also observed that when more views were provided, participants tended to leave gaps and overlaps in the design in \emph{Basic} mode, whilst they carefully aligned all views using alignment or recommendation functions in \emph{Partial} and \emph{Full} modes.
The participants complained that it is too difficult to generate a complete layout in \emph{Basic} mode.

\noindent
\textbf{Feedback}.
After the experiments, we collected feedback from the participants using 7-point Likert scale questions.
Figure~\ref{fig:design_rank} (right) presents results on the usability of the three design modes.
The participants felt that the \textit{Full} mode (mean = 6.6) is easier to use than \textit{Partial} (mean = 5.2) and \textit{Basic} mode (mean = 3.5).
Using the Bonferroni post-hoc test, we found a significant difference among the modes ($p < 0.01$).
More results are provided in Supplementary Table S5.
Overall, the participants thought the idea of example-based MV designs is promising: 
The system is novel, interesting, and useful for designing MVs (Q4); 
All participants agreed that the interface is intuitive (Q5), and interactions and recommendation results are useful (Q6);
The idea could also be useful for other applications like infographics and mobile app design (Q9).
For future work, the participants suggested showing mockup views of real data rather than icons, to consider view directions, and to export the layout as a JSON file (Q7). 
\section{Discussion, Limitation, and Future Work}
\label{sec:discussion}

We conducted a comprehensive analysis on the \emph{composition} and \emph{configuration} patterns of MVs in visualization literature.
The analyses revealed some common practices:

\begin{itemize}
\vspace{-2mm}
\item
Most MVs present less than five views. 
Designers shall be careful when the MV design compromises too many views.
Simply putting more views together may cause information overload.

\vspace{-2mm}
\item
Simpler and perceptually more accurate view types are preferred, \emph{e.g.}, \emph{Bar} and \emph{Line} charts.
On the other hand, \emph{Circle} charts, such as donut and pie charts, were seldom used.

\vspace{-2mm}
\item
Most MVs adopt simple layout, \emph{e.g.}, 2A \includegraphics[height=0.1in]{figures/tile_img/1.png} and 3A \includegraphics[height=0.1in]{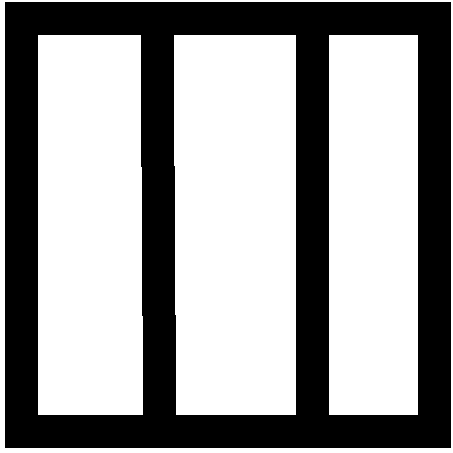}, especially for InfoVis and SciVis papers.
VAST papers tend to adopt a bit more complex layouts, as shown in Table~\ref{tab:stats}.

\vspace{-2mm}
\item
Most view types show a medium range of aspect ratio within [1/2, 2], except for some types like \emph{Area} and \emph{Panel}.

\vspace{-2mm}
\item
There are common positions for certain view types.
For instance, \emph{Panel} typically are not positioned in the center of the display space, whilst \emph{Diagrams} are the opposite.
\end{itemize}

\vspace{-2mm}

The findings advance our understanding of the design space of MVs.
The revealed composition and configuration patterns also provide a foundation for the development of a recommendation system that can assist a designer in choosing appropriate MV designs. 
Usability and effectiveness of the recommendation system are confirmed by feedbacks from both visualization novices and experts. 

\noindent
\emph{Limitations}.
Nevertheless, there are certain limitations in our work.
First, we categorize a view into one of the 14 view types.
This categorization is not exhaustive, and many views cannot be classified within any of these view types.
We found many novel designs with compound view types.
In such scenarios, we categorize the views using the main view types, \emph{e.g.}, \emph{Line} chart for\cite{yuan_2009_scattering} that combines scatter points in parallel coordinates, and \emph{Map} for\cite{zeng_2013_visualizing} that overlays a novel chord diagram on maps. 
A more solid solution would be to adopt the concept of composite visualization views\cite{javed_2012_exploring} that identified five design patterns of compositing views, \emph{i.e.}, juxtaposition, superimposition, overloading, nesting, and integration.
Second, we only analyze view type usages and view layouts, but not the underlying mechanism for selecting view types and arranging views.
There are many factors that can affect the designs, such as underlying data, analytical tasks, individual preferences, and user perceptions.
For instance, Wang et al.\cite{wang_2018_is} proposed a comprehensive strategy for determining the aspect ratio of the line chart.
A thorough consideration of all the factors would certainly be beneficial.
We aim to include these factors in the future. 
Last, the current recommendation system is solely based on MV designs from visualization conferences.
The dataset is however rather small, and the design space is limited.
For instance, few MV designs in the dataset are for large-size displays.
We will increase the corpus size by collecting more MV designs from other venues such as IEEE TVCG, ACM CHI, the Internet, and visualization authoring tools like Tableau.

\vspace{1.5mm}
\noindent
\emph{Future Work}.
There are several promising directions for future works.
First, many participants in Study 2 suggested incorporating views for real data instead of view type icons.
Such functionalities can be supported by visualization libraries like D3\cite{d3}, and we plan to realize it in the near future.
Second, we would like to analyze the underlying semantic structure that links multiple views, similar to visual information flow in infographics\cite{VIF20}.
We anticipate that studying the semantic structure would improve the visualization community's understanding on how to arrange multiple views in the display space.
Third, we would like to examine visual consistency between multiple views.
Though the visualization community proposed many guidelines in keeping visual consistency (\emph{e.g.},\cite{qu_2018_keeping}), we found many violations when annotating the MVs.
We expect to find out some criteria that can be used to evaluate MVs.
Lastly, dashboards that mostly use multiple views recently gain much attention in the visualization community.
Sarikaya et al.\cite{sarikaya_2019_what} built a dataset including a series of dashboard images and corresponding design goals.
This complements the limitation of our dataset lacking information of design goals.
We would like to analyze composition and configuration patterns of these dashboards, and associate the patterns with concrete design goals.
We anticipate that a deeper analysis would provide much clearer guidelines for MV design.
\section{Conclusion}
\label{sec:conclusion}
We have presented an empirical study on how the visualization community combines multiple views of different types in the display space, and have developed an interactive recommendation system that uses data from the analysis. 
The benefit of this work is prominent: 
First, the study advances the community’s understanding on view type associations and view layout arrangements. 
The study was conducted on the basis of a new dataset containing 360 images of MVs collected from IEEE VIS, EuroVis, and PacificVis publications 2011 to 2019, and fine-grained annotations of view types and layouts for these visualization images. 
Second, the study provides the foundation for a recommendation system that assists a designer in designing MV, by enabling faceted exploration of existing MV designs. Third the tool recommends MV designs based on revealed composition and configuration patterns and user inputs. 
We plan to release the dataset to advance future studies on MV design, and for other researchers to develop their own MV recommendation systems.
\vspace*{1.5mm}
\noindent
\acknowledgments{
The authors wish to thank all experiment participants for their valuable feedbacks, and the anonymous reviewers for their fruitful suggestions.
This work was supported in part by National Natural Science Foundation of China (No. 61802388), the National Science Foundation (OAC-1940175, OAC-1939945, IIS-1452977, DGE-1855886), and DARPA (FA8750-17-2-0107).
}

\bibliographystyle{abbrv}

\bibliography{reference}

\begin{thebibliography}{10}

\bibitem{power_bi}
{Power BI}.
\newblock \url{https://powerbi.microsoft.com}, last accessed on 26/04/2020.

\bibitem{spotfire}
{Spotfire}.
\newblock \url{https://www.tibco.com/products/tibco-spotfire/}, last accessed
  on 26/04/2020.

\bibitem{tableau}
{Tableau}.
\newblock \url{https://www.tableau.com/}, last accessed on 26/04/2020.

\bibitem{al-maneea_2019_multipleview}
H.~M. Al-maneea and J.~C. Roberts.
\newblock Towards quantifying multiple view layouts in visualisation as seen
  from research publications.
\newblock In {\em Proceedings of the IEEE Visualization}, 2019.

\bibitem{michelle_2000_guidelines}
M.~Q.~W. Baldonado, A.~Woodruff, and A.~Kuchinsky.
\newblock Guidelines for using multiple views in information visualization.
\newblock In {\em Proceedings of the Working Conference on Advanced Visual
  Interfaces}, pages 110--119, 2000.

\bibitem{bederson_2003_ordered}
B.~B. Bederson, B.~E.~N. Shneiderman, and M.~Wattenberg.
\newblock {\em Ordered and Quantum Treemaps: Making Effective Use of 2D Space
  to Display Hierarchies}, pages 257--278.
\newblock Morgan Kaufmann, 2003.

\bibitem{borkin2013makes}
M.~A. Borkin, A.~A. Vo, Z.~Bylinskii, P.~Isola, S.~Sunkavalli, A.~Oliva, and
  H.~Pfister.
\newblock What makes a visualization memorable?
\newblock {\em IEEE Transactions on Visualization and Computer Graphics},
  19(12):2306--2315, 2013.

\bibitem{d3}
M.~Bostock, V.~Ogievetsky, and J.~Heer.
\newblock {$D^3$}: Data-driven documents.
\newblock {\em IEEE Transactions on Visualization and Computer Graphics},
  17(12):2301--2309, 2011.

\bibitem{Bruls99squarifiedtreemaps}
M.~Bruls, K.~Huizing, and J.~van Wijk.
\newblock Squarified treemaps.
\newblock In {\em Proceedings of the Joint Eurographics and IEEE TCVG Symposium
  on Visualization}, pages 33--42. Press, 1999.

\bibitem{card_1999_readings}
S.~K. Card, J.~D. Mackinlay, and B.~Shneiderman.
\newblock {\em Readings in information visualization: using vision to think}.
\newblock Morgan Kaufmann, 1999.

\bibitem{casner_1991_task}
S.~M. Casner.
\newblock Task-analytic approach to the automated design of graphic
  presentations.
\newblock {\em ACM Transactions on Graphics}, 10(2):111--151, 1991.

\bibitem{chen_2019_towards}
Z.~Chen, Y.~Wang, Q.~Wang, Y.~Wang, and H.~Qu.
\newblock Towards automated infographic design: Deep learning-based
  auto-extraction of extensible timeline.
\newblock {\em IEEE Transactions on Visualization and Computer Graphics},
  26(1):917--926, 2020.

\bibitem{chung_2015_four}
H.~Chung, C.~North, S.~Joshi, and C.~Jian.
\newblock Four considerations for supporting visual analysis in display
  ecologies.
\newblock In {\em Proceedings of the IEEE Conference on Visual Analytics
  Science and Technology (VAST)}, pages 33--40, 2015.

\bibitem{cleveland_1984_graphical}
W.~S. Cleveland and R.~McGill.
\newblock Graphical perception: Theory, experimentation, and application to the
  development of graphical methods.
\newblock {\em Journal of the American Statistical Association},
  79(387):531--554, 1984.

\bibitem{collins_2007_vislink}
C.~Collins and S.~Carpendale.
\newblock {VisLink}: Revealing relationships amongst visualizations.
\newblock {\em IEEE Transactions on Visualization and Computer Graphics},
  13(6):1192--1199, 2007.

\bibitem{dibia_2019_data2vis}
V.~Dibia and D.~C.
\newblock {Data2Vis}: Automatic generation of data visualizations using
  sequence-to-sequence recurrent neural networks.
\newblock {\em IEEE Computer Graphics and Applications}, 39(5):33--46, 2019.

\bibitem{gleicher_2011_visual}
M.~Gleicher, D.~Albers, R.~Walker, I.~Jusufi, C.~D. Hansen, and J.~C. Roberts.
\newblock Visual comparison for information visualization.
\newblock {\em Information Visualization}, 10(4):289--309, 2011.

\bibitem{horak_2019_vistribute}
T.~Horak, A.~Mathisen, C.~N. Klokmose, R.~Dachselt, and N.~Elmqvist.
\newblock Vistribute: Distributing interactive visualizations in dynamic
  multi-device setups.
\newblock In {\em Proceedings of the ACM Conference on Human Factors in
  Computing Systems}, pages 616: 1--13, 2019.

\bibitem{hu_2019_vizml}
K.~Hu, M.~A. Bakker, S.~Li, T.~Kraska, and C.~Hidalgo.
\newblock {VizML}: A machine learning approach to visualization recommendation.
\newblock In {\em Proceedings of the ACM Conference on Human Factors in
  Computing Systems}, pages 128: 1--12, 2019.

\bibitem{hu_2019_viznet}
K.~Hu, S.~N.~S. Gaikwad, M.~Hulsebos, M.~A. Bakker, E.~Zgraggen, C.~Hidalgo,
  T.~Kraska, G.~Li, A.~Satyanarayan, and C.~Demiralp.
\newblock {VizNet}: Towards a large-scale visualization learning and
  benchmarking repository.
\newblock In {\em Proceedings of the ACM Conference on Human Factors in
  Computing Systems}, pages 662: 1--12, 2019.

\bibitem{isenberg_2017_vispubdata}
P.~Isenberg, F.~Heimerl, S.~Koch, T.~Isenberg, P.~Xu, C.~D. Stolper,
  M.~Sedlmair, J.~Chen, T.~Möller, and J.~Stasko.
\newblock {vispubdata.org}: A metadata collection about {IEEE Visualization
  (VIS)} publications.
\newblock {\em IEEE Transactions on Visualization and Computer Graphics},
  23(9):2199--2206, 2017.

\bibitem{javed_2012_exploring}
W.~Javed and N.~Elmqvist.
\newblock Exploring the design space of composite visualization.
\newblock In {\em Proceedings of the IEEE Pacific Visualization Symposium},
  pages 1--8, 2012.

\bibitem{javed_2013_explates}
W.~Javed and N.~Elmqvist.
\newblock {ExPlates}: Spatializing interactive analysis to scaffold visual
  exploration.
\newblock {\em Computer Graphics Forum}, 32(3pt4):441--450, 2013.

\bibitem{johnson_1991_tree-maps}
B.~Johnson and B.~Shneiderman.
\newblock Tree-maps: a space-filling approach to the visualization of
  hierarchical information structures.
\newblock In {\em Proceeding of IEEE Visualization}, pages 284--291, 1991.

\bibitem{alicia_2012_vizdeck}
A.~Key, B.~Howe, D.~Perry, and C.~Aragon.
\newblock {VizDeck}: self-organizing dashboards for visual analytics.
\newblock In {\em Proceedings of the ACM SIGMOD International Conference on
  Management of Data}, pages 681--684, 2012.

\bibitem{langner_2018_vistiles}
R.~Langner, T.~Horak, and R.~Dachselt.
\newblock {VisTiles}: Coordinating and combining co-located mobile devices for
  visual data exploration.
\newblock {\em IEEE Transactions on Visualization and Computer Graphics},
  24(1):626--636, 2018.

\bibitem{langner_2019_multiple}
R.~Langner, U.~Kister, and R.~Dachselt.
\newblock Multiple coordinated views at large displays for multiple users:
  Empirical findings on user behavior, movements, and distances.
\newblock {\em IEEE Transactions on Visualization and Computer Graphics},
  25(1):608--618, 2019.

\bibitem{gerald_1994_classification}
G.~L. Lohse, K.~Biolsi, N.~Walker, and H.~H. Rueter.
\newblock A classification of visual representations.
\newblock {\em Communication of the ACM}, 37(12):36--49, 1994.

\bibitem{VIF20}
M.~Lu, C.~Wang, J.~Lanir, N.~Zhao, H.~Pfister, D.~Cohen-Or, and H.~Huang.
\newblock Exploring visual information flows in infographics.
\newblock In {\em Proceedings of the ACM Conference on Human Factors in
  Computing Systems}, pages 1--12, 2020.

\bibitem{luo_2018_deepeye}
Y.~Luo, X.~Qin, N.~Tang, and G.~Li.
\newblock {DeepEye}: Towards automatic data visualization.
\newblock In {\em Proceedings of the IEEE International Conference on Data
  Engineering}, pages 101--112, 2018.

\bibitem{mackinlay_1986_automating}
J.~Mackinlay.
\newblock Automating the design of graphical presentations of relational
  information.
\newblock {\em ACM Transactions on Graphics}, 5(2):110--141, 1986.

\bibitem{mackinlay_2007_tabulea}
J.~Mackinlay, P.~Hanrahan, and C.~Stolte.
\newblock Show me: Automatic presentation for visual analysis.
\newblock {\em IEEE Transactions on Visualization and Computer Graphics},
  13(6):1137--1144, 2007.

\bibitem{matkovic_2008_comvis}
K.~Matkovic, W.~Freiler, D.~Gracanin, and H.~Hauser.
\newblock {ComVis}: A coordinated multiple views system for prototyping new
  visualization technology.
\newblock In {\em Proceedings of the International Conference Information
  Visualisation}, pages 215--220, 2008.

\bibitem{moritz_2019_draco}
D.~Moritz, C.~Wang, G.~L. Nelson, H.~Lin, A.~M. Smith, B.~Howe, and J.~Heer.
\newblock Formalizing visualization design knowledge as constraints: Actionable
  and extensible models in {Draco}.
\newblock {\em IEEE Transactions on Visualization and Computer Graphics},
  25(1):438--448, 2019.

\bibitem{donovan_2014_learning}
P.~O’Donovan, A.~Agarwala, and A.~Hertzmann.
\newblock Learning layouts for single page graphic designs.
\newblock {\em IEEE Transactions on Visualization and Computer Graphics},
  20(8):1200--1213, 2014.

\bibitem{qu_2018_keeping}
Z.~Qu and J.~Hullman.
\newblock Keeping multiple views consistent: Constraints, validations, and
  exceptions in visualization authoring.
\newblock {\em IEEE Transactions on Visualization and Computer Graphics},
  24(1):468--477, 2018.

\bibitem{yolov3}
J.~Redmon and A.~Farhadi.
\newblock {YOLOv3}: An incremental improvement.
\newblock {\em ArXiv e-prints}, 2018.

\bibitem{roberts_1998_encouraging}
J.~C. Roberts.
\newblock On encouraging multiple views for visualization.
\newblock In {\em Proceedings of the IEEE Conference on Information
  Visualization.}, pages 8--14, 1998.

\bibitem{roberts_2007_state}
J.~C. Roberts.
\newblock State of the art: Coordinated \& multiple views in exploratory
  visualization.
\newblock In {\em Proceedings of International Conference on Coordinated and
  Multiple Views in Exploratory Visualization}, pages 61--71, 2007.

\bibitem{roberts_2019_multiple}
J.~C. Roberts, H.~Al-maneea, P.~W.~S. Butcher, R.~Lew, G.~Rees, N.~Sharma, and
  A.~Frankenberg-Garcia.
\newblock Multiple views: different meanings and collocated words.
\newblock {\em Computer Graphics Forum}, 38(3):79--93, 2019.

\bibitem{sadana_2016_design}
R.~Sadana and J.~Stasko.
\newblock Designing multiple coordinated visualizations for tablets.
\newblock {\em Computer Graphics Forum}, 35(3):261--270, 2016.

\bibitem{saket_2018_beyond}
B.~Saket, D.~Moritz, H.~Lin, V.~Dibia, C.~Demiralp, and J.~Heer.
\newblock Beyond heuristics: Learning visualization design.
\newblock {\em arXiv:1807.06641}, pages 1--4, 2018.

\bibitem{sarikaya_2019_what}
A.~Sarikaya, M.~Correll, L.~Bartram, M.~Tory, and D.~Fisher.
\newblock What do we talk about when we talk about dashboards?
\newblock {\em IEEE Transactions on Visualization and Computer Graphics},
  25(1):682--692, 2019.

\bibitem{shneiderman_1992_tree}
B.~Shneiderman.
\newblock Tree visualization with tree-maps: {2-d} space-filling approach.
\newblock {\em ACM Transactions on Graphics}, 11(1):92--99, 1992.

\bibitem{schneiderman_1996_eyes}
B.~Shneiderman.
\newblock The eyes have it: a task by data type taxonomy for information
  visualizations.
\newblock In {\em Proceedings of the IEEE Symposium on Visual Languages}, pages
  336 -- 343, 1996.

\bibitem{sondag_2018_stable}
M.~Sondag, B.~Speckmann, and K.~Verbeek.
\newblock Stable treemaps via local moves.
\newblock {\em IEEE Transactions on Visualization and Computer Graphics},
  24(1):729--738, 2018.

\bibitem{stolte_2002_polaris}
C.~Stolte, D.~Tang, and P.~Hanrahan.
\newblock Polaris: a system for query, analysis, and visualization of
  multidimensional relational databases.
\newblock {\em IEEE Transactions on Visualization and Computer Graphics},
  8(1):52--65, 2002.

\bibitem{swearngin_2018_rewire}
A.~Swearngin, M.~Dontcheva, W.~Li, J.~Brandt, M.~Dixon, and A.~J. Ko.
\newblock Rewire: Interface design assistance from examples.
\newblock In {\em Proceedings of the ACM Conference on Human Factors in
  Computing Systems}, pages 1--12, 2018.

\bibitem{tufte_1983_visual}
E.~R. Tufte.
\newblock {\em The Visual Display of Quantitative Information}.
\newblock Graphics Press, second edition, 1983.

\bibitem{vartak_2015_seedb}
M.~Vartak, S.~Rahman, S.~Madden, A.~Parameswaran, and N.~Polyzotis.
\newblock Seedb: Efficient data-driven visualization recommendations to support
  visual analytics.
\newblock {\em Proceedings of the VLDB Endowment.}, 8(13):2182--2193, 2015.

\bibitem{wang_2018_is}
Y.~Wang, Z.~Wang, L.~Zhu, J.~Zhang, C.~Fu, Z.~Cheng, C.~Tu, and B.~Chen.
\newblock Is there a robust technique for selecting aspect ratios in line
  charts?
\newblock {\em IEEE Transactions on Visualization and Computer Graphics},
  24(12):3096--3110, 2018.

\bibitem{weaver_2004_improvise}
C.~Weaver.
\newblock Building highly-coordinated visualizations in improvise.
\newblock In {\em Proceedings of the IEEE Symposium on Information
  Visualization}, pages 159--166, 2004.

\bibitem{wu_2018_miqp}
W.~Wu, L.~Fan, L.~Liu, and P.~Wonka.
\newblock Miqp-based layout design for building interiors.
\newblock {\em Computer Graphics Forum}, 37(2):511--521, 2018.

\bibitem{Wu_2016_TelCoVis}
W.~{Wu}, J.~{Xu}, H.~{Zeng}, Y.~{Zheng}, H.~{Qu}, B.~{Ni}, M.~{Yuan}, and L.~M.
  {Ni}.
\newblock Telcovis: Visual exploration of co-occurrence in urban human mobility
  based on telco data.
\newblock {\em IEEE Transactions on Visualization and Computer Graphics},
  22(1):935--944, 2016.

\bibitem{xu_2014_global}
P.~Xu, H.~Fu, T.~Igarashi, and C.-L. Tai.
\newblock Global beautification of layouts with interactive ambiguity
  resolution.
\newblock In {\em Proceedings of the ACM symposium on User interface software
  and technology}, pages 243--252, 2014.

\bibitem{yang_2013_urban}
Y.-L. Yang, J.~Wang, E.~Vouga, and P.~Wonka.
\newblock Urban pattern: layout design by hierarchical domain splitting.
\newblock {\em ACM Transactions on Graphics}, 32(6):1--12, 2013.

\bibitem{yuan_2009_scattering}
X.~Yuan, P.~Guo, H.~Xiao, H.~Zhou, and H.~Qu.
\newblock Scattering points in parallel coordinates.
\newblock {\em IEEE Transactions on Visualization and Computer Graphics},
  15(6):1001--1008, 2009.

\bibitem{zeng_2013_visualizing}
W.~Zeng, C.-W. Fu, S.~M{\"u}ller~Arisona, and H.~Qu.
\newblock Visualizing interchange patterns in massive movement data.
\newblock {\em Computer Graphics Forum}, 32(3pt3):271--280, 2013.

\bibitem{zheng_2019_content-aware}
X.~Zheng, X.~Qiao, Y.~Cao, and R.~W.~H. Lau.
\newblock Content-aware generative modeling of graphic design layouts.
\newblock {\em ACM Transactions on Graphics}, 38(4):1--15, 2019.

\end{thebibliography}
\end{document}